\documentclass[10pt,journal,compsoc]{IEEEtran}

\ifCLASSINFOpdf
\else
 \fi

\usepackage[ruled,vlined,linesnumbered]{algorithm2e}
\usepackage{amsthm}
\usepackage{amsmath}
\usepackage{graphicx}
\usepackage[compatible]{algpseudocode}
\usepackage{amssymb}
\usepackage{amsfonts}
\usepackage{algcompatible}
\usepackage{tabularx}
\usepackage{lipsum}

\DeclareMathOperator*{\argmax}{arg\,max}

\usepackage[utf8]{inputenc}
\newtheorem{theorem}{Theorem}

\usepackage{url}
\usepackage{xcolor}
\usepackage{balance}
\usepackage{subcaption}
\usepackage{epstopdf}
\usepackage{array}
\usepackage{comment}
\usepackage[noadjust]{cite}
\usepackage{amsthm}
\usepackage{tipa}
\usepackage{breqn}

\newtheorem{lemma}{Lemma}
\SetKw{Continue}{continue}
\SetKw{Break}{break}
\SetKwFor{RepTimes}{repeat}{times}{end}

\begin{document}
\title{Criticality and Utility-aware Fog Computing System for Remote Health Monitoring}
\author{Moirangthem Biken Singh, Navneet Taunk, Naveen Kumar Mall, and  Ajay~Pratap,~\IEEEmembership{Member,~IEEE}
	\IEEEcompsocitemizethanks{\IEEEcompsocthanksitem M. B. Singh, N. Taunk, N. K. Mall and A. Pratap are with the Department of Computer Science and Engineering, Indian Institute of Technology (Banaras Hindu University) Varanasi 221005 India. E-mail: \{moirangthembsingh.rs.cse21, navneettaunk.cse18, naveenkumarmall.cse18, ajay.cse\}@iitbhu.ac.in.\protect 
    	       }
	}

\IEEEtitleabstractindextext{
\begin{abstract}
Growing remote health monitoring system allows constant monitoring of the patient's condition and performance of preventive and control check-ups outside medical facilities. However, the real-time smart-healthcare application poses a delay constraint that has to be solved efficiently. Fog computing is emerging as an efficient solution for such real-time applications. Moreover, different medical centers are getting attracted to the growing IoT-based remote healthcare system in order to make a profit by hiring Fog computing resources. However, there is a need for an efficient algorithmic model for allocation of limited fog computing resources in the criticality-aware smart-healthcare system considering the profit of medical centers. Thus, the objective of this work is to maximize the system utility calculated as a linear combination of the profit of the medical center and the loss of patients. To measure profit, we propose a flat-pricing-based model. Further, we propose a swapping-based heuristic to maximize the system utility. The proposed heuristic is tested on various parameters and shown to perform close to the optimal with criticality-awareness in its core. Through extensive simulations, we show that the proposed heuristic achieves an average utility of $96\%$ of the optimal, in polynomial time complexity.

\end{abstract}
}

\maketitle

\IEEEdisplaynontitleabstractindextext

\IEEEpeerreviewmaketitle
   
\section{Introduction}\label{Sec1}

IoT assisted remote healthcare has recently gained popularity as it seemed to be an efficient solution for the challenges faced in remote healthcare sector \cite{9319251}. The lack of healthcare facilities in rural \emph{India} can be greatly assisted by IoT-based remote health monitoring in a cost effective way \cite{dash2020impact}. However, the development of IoT tackles various problems such as limitation of available resources, limited accessibility of IoT systems for people living in rural areas.

In remote health monitoring, the patient is equipped with sensors, and the data generated by the sensor is sent to a gateway through Wireless Body Area Network (WBAN). Further, the gateway device sends the data to the base station through beyond-WBAN. In beyond-WBANs, 5G communication is emerging an efficient solution for fast and real-time transmissions \cite{9371426, bishoyi2021enabling}. The assistance of fog computing within beyond-WBAN has emerged as an efficient way to compute the data sent through the network \cite{mutlag2019enabling}. Although their computational capacities are not as much as centralized cloud servers, they are capable enough of computing medical data packets near to patients \cite{yu2017survey}. Therefore, fog computing can reduce the latency to a greater extent, thus improving the quality of the monitoring system \cite{yi2015survey}.

The people living in rural areas have a higher rate of poverty \cite{deshingkar2010migration} and thus, they cannot afford Local Devices (LDs) or sensors on their own. They need to be assisted by the government with these pieces of equipment. Thus, a better solution is to provide these things for free or at low price. Moreover, the charge for monitoring should be kept low. Therefore, the profit of the medical center becomes an important factor for the system model as the revenue generated would be less due to cheap monitoring cost. The medical centers providing remote healthcare to these patients would see the technological advancement for their benefit and thus, they would invest in it to make a profit out of it.

In this work, the proposed system is divided into two parts, one is intra-WBAN and the other is beyond-WBAN. Intra-WBAN consists of sensors deployed on the patients and the LD provided to the patients whereas beyond-WBAN consists of different LDs that send the data to the Fog Servers (FSs) located near the base station/access point. The encouragement of such model is described as below: 

\subsection{Motivation} 
The emerging FS assisted remote health monitoring system can be fruitful for both the patients and the medical centers as discussed below:
\begin{itemize}
    \item As the medical data is highly critical, it has to be processed in real-time without much delay.
    \item The data of patients with a serious disease or other critical conditions should be given higher priority over others.
    \item In a practical scenario, a medical center would charge the patients for the medical service it provides.
    \item Therefore, there is a need of a well established cost effective mechanism between patient and healthcare service provider to monitor criticality, latency and revenue via resource allocation in remote healthcare system.   
\end{itemize}

Motivated by the above objectives, we aim to formulate an FS assisted beyond-WBAN based remote health monitoring system that minimizes the cost of patients with the profit of the medical center in consideration. Inspired by \cite{ning2020mobile}, we aim to use dedicated LDs to not only collect the data sent by the sensors but also those LDs have some computation power, which can be utilized to compute the patients' data locally if the condition of the patient is not much critical.  

\subsection{Contribution}
In this paper, we design an FS assisted remote health monitoring system. The main objective of the system is to maximize the utility that depends on the profit of the medical center and the cost of patients (which depends on latency delay and their criticality described in Section \ref{Sec3}). The main contributions of this paper are summarized below:
\begin{itemize}
    \item A criticality and utility-aware remote health monitoring system is proposed.
    \item A cost-function is formulated for the patients that measure the loss of patients in terms of latency and their criticality in such a way that more critical patients are given priority over less critical patients. Moreover, the  problem is formulated depending on the patients' loss and the medical center's profit, trying to balance between the two.
    \item A swapping-based heuristic is proposed to maximize the system utility under the constraints of permissible latency for the computation of patients' data in polynomial time complexity.
    \item Through extensive simulations, the proposed heuristic is found to achieve a utility of $96\%$ of the optimal on an average.
    
\end{itemize}

The rest of the paper is organized as follows: Section \ref{SecII} reviews the relevant work. The system model and the problem definition are introduced in Sections \ref{Sec3} and \ref{Sec4}, respectively. The proposed solution and analysis are given in Sections \ref{Sec6} and \ref{analysis}, respectively. The performance study is presented in Section \ref{Sec8}. Finally, Section \ref{Sec9} offers conclusions and future research directions. 

\section{Related Works}\label{SecII}

The authors in \cite{feng2019optimal} propose a haptic communication framework for e-health systems. The primary focus of the paper is to improve haptic communications under three factors (system stability, energy consumption, and network delay). They propose a time-varying swarm algorithm to solve the formulated problem. The authors in \cite{gu2015cost} propose a cost-aware medical cyber-physical system assisted by fog computing. Their work jointly focuses on task allocation, base station association, and virtual machine placement. They propose linear-programming based heuristics to solve the formulated problem. The authors in \cite{apat2020energy} propose an energy-aware medical cyber-physical system assisted by fog computing. Their primary focus is on resource allocation to minimize energy consumption and response time. They propose a dynamic-cluster algorithm to solve the formulated problem.

In \cite{qiu2021computation}, the authors investigate the factors such as energy consumption, transmission delay, QoS requirement, the power limit and wireless fronthaul constraint in fog computing-based Internet of Medical Things (IoMT) for remote health monitoring. They propose a low time-complexity sub-optimal scheme to solve the problem. The authors in \cite{yi2018transmission} propose a queue-based transmission of time-sensitive medical data packets in beyond-WBAN. They propose a non-cooperative game for the above mentioned scenarios and then, they propose an analytical framework to solve the formulated problem. The authors in \cite{ning2020mobile} propose a health monitoring system for IoMT considering criticality, energy and delay constraints. They propose a decentralized non-cooperative game based scheme to solve the formulated problem.  

The authors in \cite{misra2014priority} propose a priority-aware time-slot allocation in WBANs. They extend the evolutionary game theory to solve the formulated problem. The authors in \cite{misra2015cooperative} propose a Nash bargaining solution for a cooperative game based priority-aware data-rate tuning in WBANs model. Moreover, Table \ref{related_works} summarizes the closely related works available in the literature.

\textbf{Shortcomings  of  Existing  Approaches}: 
In some of the existing approaches, only intra-WBAN transmission is considered on the basis of latency and criticality. Some works consider both intra-WBAN and beyond-WBAN transmission under latency and criticality constraints. However, none of the existing approach have considered the profit of the medical center which is one of the main objective of our system. Therefore, different from the above work, we propose a novel criticality-aware health monitoring system with the profit of the medical center in consideration. Moreover, we propose a novel swapping-based heuristic to solve the formulated problem in polynomial time complexity.

\begin{table}[h!]
    \centering
     \caption{A relative comparison} \label{related_works}
    \begin{tabular}{|p{1.4cm}||p{2.5cm}||p{2.5cm}|}
       \hline
       Author & Primary Problem & Brief Description\\
       \hline
       Feng et. al. \cite{feng2019optimal}& Resource Allocation, Packet Drop, Energy Harvesting & Swarm Intelligence\\
       \hline
       Gu et. al. \cite{gu2015cost}& Task Allocation, Base Station Association, Machine Placement & LP-based heuristics\\
       \hline
       Apat et. al. \cite{apat2020energy}& Energy Consumption, Response Time, Resource Allocation & Dynamic clustering\\
       \hline
       Qui et. al. \cite{qiu2021computation}& QoS requirement, power limit and wireless fronthaul constraint. & Lagrange Multipliers based\\
       \hline
       Yi et. al. \cite{yi2018transmission}& Data Priority, Latency & Non-cooperative Game\\
       \hline
       Ning et. al. \cite{ning2020mobile}& Medical Criticality, Age Of Information, Energy Consumption & Non-cooperative Game\\
       \hline
       Misra et. al. \cite{misra2014priority}& Data Priority, Time-slot Allocation & Evolutionary Game Theory\\
       \hline
       Misra et. al. \cite{misra2015cooperative}& Data Priority, Data-rate Tuning & Cooperative Bargaining Game\\
       \hline
       Proposed model& Critcality, Latency, Profit and Resource Allocation & Swapping-based heuristic \\
       \hline
       
    \end{tabular}
  \end{table}

\section{System Model}\label{Sec3}

\begin{table}[h!]
	\begin{center}
	\vspace{-5pt}
		\caption{Symbol description}
		\label{SybDes}
		\begin{tabular}{ |p{1.3cm}||p{5cm}|  }
			\hline
			\textbf{Symbol} & \textbf{Description}\\
			
			\hline
			$\mathbb{S}$ & Set of sensors for a patient \\
			$\mathbb{X}$ & Set of medical criticalites for all sensors.\\
			$x_{s}$ &  Medical Criticality of sensor $s$.\\
			$\theta_{s,t}$ & Physiological data value sensed by the sensor $s$ at time $t$\\
			 $\theta_{l,s}$ & Lower limit of the normal value for sensor $s$.\\
			$\theta_{u,s}$ & Upper limit of the normal value for sensor $s$. \\
			$d_{s,t}$ & Packet severity index for sensor $s$ at time $t$. \\ 
			$c_{s,t}$ & Overall criticality index for sensor $s$ at time $t$. \\
			$\mathbb{F}$ & Set of FSs.\\
			$\mathbb{P}$ & Set of patients.\\
			$\rho_{p,t}^{c}$ & Patient Criticality for patient $p$ at time $t$.\\
			$\mathbb{H}_t$ & Set of strategy for patients.\\
			$u_{p,t}$ & Whether LD is chosen for computation\\
			$q_{p,t}$ & Whether FS is chosen for computation\\
			$\eta_{p,t}$ & Overall data size for patient $p$ at time $t$.\\
			$\beta_{p,t}$ & CPU cycles required to compute patient $p$'s data at time $t$.\\
			$T_{p,t}^{c,l}$ & Computation time for patient $p$ at time $t$ by local device.\\
			$\Upsilon$ & Computation capacity of patient's local device. \\
			$\Gamma$ & Computation capacity of an edge server. \\ 
			$m$ & Price per unit time for computation at FS\\
			$l$ & Price per unit time for computation at LD\\
			$\chi_{t}$ & Revenue earned by medical center. \\
			$\phi_{t}$ & Expenditure of medical center per edge server.\\
			$g$ & Expenditure of medical center per CPU cycle of computation at edge server.\\
			$k$ & Fixed charge per FS.\\
			$\delta$ & Latency constraint.\\
			
		\hline
		\end{tabular}
	\end{center}
	\vspace{-15pt}
\end{table}

As shown in Fig. \ref{model}, we consider a remote health monitoring system, provided by medical center to a set of patients $\mathbb{P} = \{1,2,..,P\}$\footnote{The symbol description is given in the Table \ref{SybDes}.}. The proposed problem setting is equivalent to project assignment problem in colleges where a student approaches to a professor for project assignment and professor assigns an available project to the student. Similarly in our case, patient (LD) approaches to medical center for remote monitoring and medical center assigns an available FS to the patient (LD) as shown in Fig. \ref{model}. 

The proposed problem setting is divided into two parts-intra-WBAN and beyond-WBAN described as follows:

\begin{figure}[h!]
	\centering
% 	\vspace{-15pt}
	\includegraphics[height=6cm, width=8.0cm]{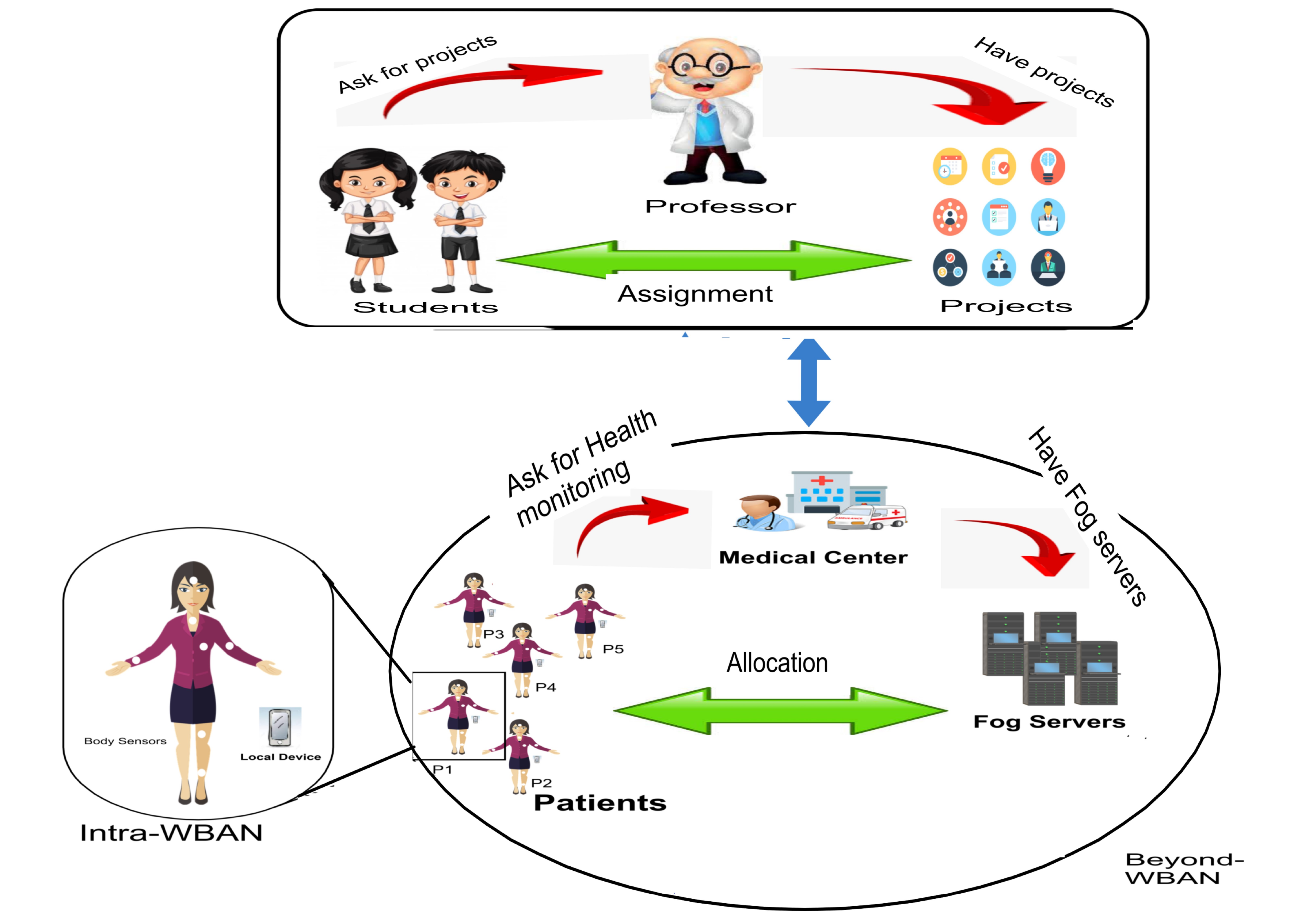}
	\caption{System Model}
	\label{model}
% 	\vspace{-15pt}
\end{figure}

\subsection{Intra-WBAN}

Consider a set of sensors, $\mathbb{S} = \{1,2,..,S\}$, deployed on each patient. Each sensor generates data packets and transmits them to the LD. The data packets generated by different sensors belong to different classes depending on their criticality. For instance, a sensor measuring heart rate should have higher priority over a sensor that measures skin temperature. Moreover, heart-related diseases are more serious and should be prioritized over general diseases. To facilitate this, we have considered medical criticality as a measure to prioritise the data packet.

Let medical criticalities of data generated by sensors be the set $\mathbb{X} = \{x_1,x_2,..x_S\}$. If data generated by sensor $s$ is more critical than that of sensor $s'$, then $x_{s} > x_{s'}$. It can be possible for two sensors of the same criticality class to have different medical criticalities. Here, $x_s \in [0,\infty]$. Let $\theta_{s,t}$ be the the parameter value sensed by sensor $s$ at time $t$. Let $\theta_{l,s}$ and $\theta_{u,s}$ be the reference range of that parameter  under normal conditions for a healthy person. Then, packet severity index \cite{misra2014priority} at time $t$ can be defined as,
\begin{equation}\label{deviation}
   d_{s,t} = \left|\frac{(\theta_{u,s}-\theta_{s,t})^2 - (\theta_{s,t}-\theta_{l,s})^2}{(|\theta_{u,s}| + |\theta_{l,s}|)^2} \right|
\end{equation}
Let overall criticality index, $c_{s,t}$ of a sensor $s$ at time $t$ be the product of packet severity index and the medical criticality as follows:
\begin{equation}\label{OverallCriticality}
  c_{s,t} = x_{s}d_{s,t}.  
\end{equation}

Let $p^{th}$ patient's criticality at time $t$ be defined as the summation of overall criticality indices of sensors, as follows:
\begin{equation}\label{PatientCriticality}
   \rho^{c}_{p,t} = \sum_{s=1}^{S}c_{p,s,t}, 
\end{equation}
where $c_{p,s,t}$ is the overall criticality index for patient $p$ and sensor $s$ at time $t$. The patient criticality indicates the extent to which the health of a patient is critical. Higher the criticality value indicates more severe is the condition of the patient. After collecting the data at LD, there is a need to make a decision for its computation either at LD or FS in-order to achieve the system's constraints. Moreover, computation capacity of all LDs are considered to be uniform and equal to $\Upsilon$. Let $u_{p,t}$ be a binary variable defined as below:
\begin{equation}\label{LocalComp}
u_{p,t} =
\begin{cases}
1, & \text{System selects LD for computation of $p$'s data};
\\[1ex]
0, & \text{otherwise}.
\end{cases}
\end{equation}
Moreover, computation time at LD for a patient $p$ can be calculated as:
\begin{equation}\label{CompTimeLocal}
    T^{c,l}_{p,t} = \frac{\beta_{p,t}}{\Upsilon}
\end{equation}
where $\beta_{p,t}$ is the number of CPU cycles required for the computation of patient $p$'s data at time $t$. In the next section, we describe the transmission\footnote{Transmission of data in intra-WBAN is beyond the scope of this work. However, the existing approach \cite{ning2020mobile}, can be utilized for intra-WBAN communication.} and computation of data at FSs.

 \subsection{Beyond-WBAN}
 In this model, transmission and computation latencies of patient's data are evaluated. Consider a set of FSs $\mathbb{F} = \{ 1,2,3,..,F\}$ and let $q_{p,t}$ be a binary variable defined as follows:
\begin{equation}\label{FogComp}
q_{p,t} = 1- u_{p,t}. 
\end{equation}

Let $\mathbb{H}_{t}$ be a $P \times F$ binary matrix which denotes the choice of the system for computation of patients' data at time $t$ where, $\mathbb{H}_{t}$ is a globally accessible variable maintained at cloud server responsible for execution of the proposed heuristic discussed in the Section \ref{Sec6}. 

\begin{equation}
    \mathbb{H}_{t} = 
    \begin{bmatrix}
    h_{1,t}^{1} & h_{1,t}^{2} & . & . & h_{1,t}^{F}\\
    h_{2,t}^{1} & . & . & . & .\\
    . & . & . & . & .\\
    . & . & . & . & .\\
    h_{P,t}^{1} & . & . & . & h_{P,t}^{F}
    \end{bmatrix}
\end{equation}
where, \begin{equation}\label{Strategy}
h_{p,t}^f =
\begin{cases}
1, & \text{FS $f$ computes patient $p$'s data};
\\[1ex]
0, & \text{otherwise}.
\end{cases}
\end{equation}

The transmission rate between patient $p$ and FS $f$ can be computed as follows: 
\begin{equation}\label{BitrateBeyond}
  BR_{p,f,t} = \Omega \log_{2}(1 + \frac{W^{bey}_{p,f,t}}{N^{bey}_{t}}),
\end{equation}
where, $\Omega$ is the channel bandwidth, $W_{p,f,t}^{bey}$ be the transmission power, and $N_{t}^{bey}$ be the noise power. Mathematical expression for $W_{p,f,t}^{bey}$ can be written as,
$W_{p,f,t}^{bey} = w_{p}G_{p,f,t}$, where $w_{p}$ is the power of transmission and $G_{p,f,t}$ is the channel gain of patient $p$, if $f$ server is chosen. Here, we have assumed that all patients communicate through different channels, so interference is not considered\footnote{However, interference can be solved by applying methods given in \cite{9371426}, \cite{di2016sub}.}. Moreover, transmission time between patient $p$ and FS $f$ can be written as:
\begin{equation}\label{TimeBeyond}
    T_{p,f,t}^{tr} = \frac{\eta_{p,t}}{BR_{p,f,t}},
\end{equation}
where, $\eta_{p,t}$ is the size of the patient $p$'s data. Let computation capacity of an FS be $\Gamma$. Similar to the studies \cite{pratap2019maximizing}, \cite{pratap2020bandwidth} we are assuming that all the patients use the resource of an FS equally and all FS have the same computation capacity. The computation time for a patient $p$ due to fog computing can be calculated as:
\begin{equation}\label{TimeFog}
    T^{c,f}_{p,t}(\mathbb{H}_{t}) = \frac{\beta_{p,t}}{\gamma_{p}(\mathbb{H}_{t})},
\end{equation}
where $\gamma_{p}(\mathbb{H}_{t})$ is the fraction of resource used by the patient $p$ which can be computed by dividing total resources of a server by the number of patients utilizing it. Mathematically,
\begin{equation}\label{FogDivision}
    \gamma_{p}(\mathbb{H}_{t}) = \frac{\Gamma}{\sum_{f = 1}^{F}h_{p,t}^{f}\sum_{p' = 1}^{P}h_{p',t}^{f}}.
\end{equation}

%%%%%%%%%%%%%%%%%%%%
% Revision
%%%%%%%%%%%%%%%%%%%%

The criticality-awareness with low-latency is an important factor for the system. Thus, one of our objectives is to minimize the loss incurred for the patients which is a parameter of the criticality-awareness. The cost function for the patients can be computed as the weighted sum of computation time and the transmission time, where the weights are the patients' criticality. Mathematically,

\begin{dmath}\label{PatientCost}
      J(t) = \sum_{p \in \mathbb{P}} \rho_{p,t-1}^{c} \left(\sum_{f=1}^{F} h_{p,t}^{f} T_{p,f,t}^{tr}
      +  T_{p,t}^{c,f}(\mathbb{H}_{t}) + u_{p,t} T_{p,t}^{c,l}\right).
\end{dmath}
We can see that the cost function depends on the criticality of patients, i.e., if a patient is more critical, it will add up more to the cost, thus we have to lower the latency for that patient in the beyond-WBAN in order to reduce the cost.

As mentioned in Fig. \ref{model}, a medical center is considered for providing remote healthcare to the patients in return of service charges. As the flat-type pricing scheme is emerging as a good business model for the health providers \cite{tanwar2020optimal}, we are considering a flat-type pricing scheme to calculate the revenue of the medical center in the following.

\subsubsection{Flat-type Pricing Scheme}
For the computation on the LD, the medical center will charge $l$ unit price per time slot. If the computation is done on the FS, the medical center will charge $m$ unit price per time slot. As FS charge should be greater than that of LD, thus $m > l$. So, the revenue earned by the medical center can be calculated as, 
\begin{equation}\label{revenueI}
    \chi_{t} = \sum_{p \in \mathbb{P}}(u_{p,t}l + q_{p,t}m).
\end{equation}

Now, as FSs are involved, the medical center has to bear its cost as well. Let $k$ be the fixed expenditure per FS per time slot and $g$ be the expenses of medical center per CPU cycle due to computation on the FS. Thus, the expenditure of medical center can be calculated as,
\begin{equation}\label{expenditureFirst}
\phi_{t} = kF + g\sum_{p \in \mathbb{P}}q_{p,t}\beta_{p,t}.  
\end{equation}
Therefore, the profit gained by the medical center can be calculated as,
\begin{equation}\label{profitI}
    \Delta(t) = \chi_{t} - \phi_{t} = \sum_{p \in \mathbb{P}}(u_{p,t}l + q_{p,t}m) - kF - g\sum_{p \in \mathbb{P}}q_{p,t}\beta_{p,t}. 
\end{equation} 
On observing equation \eqref{profitI}, we can see that, for a patient $p$, the profit depends on whether that patient is allocated an FS or LD. Let the maximum value of $\beta_{p,t}$ be $\beta_{p,t}^{max}$. Here, $\beta_{p,t}^{max}$ can be approximated by the medical center before deciding the values of $m$ and $l$. The profit earned by the medical center depends on the following constraint:

\begin{equation}\label{AdditionalConstraint}
   m - l \geq g\beta_{p,t}^{max} + \frac{kF}{P}
\end{equation}
Here, it is ensured that if a patient is using an FS, then the profit of the medical center will be more compared to the case when he uses an LD independent of the CPU cycles of the patient data as discussed in Lemma \ref{Claim}. 
\begin{lemma}\label{Claim}
    The profit of the medical center either increases or remains constant as more patients utilize FS rather than LD for their computation.
\end{lemma}

\begin{proof}
    Let $P'$ be the number of patients utilizing FS. So, the profit in this scenario can be given by:
    \begin{equation}
        \Delta_{t,1} = P'm + (P - P')l - kF - g\sum_{p \in P}q_{p,t}\beta_{p,t}.
    \end{equation}
    Now, take any patient $p'$ that is utilizing LD and assign him any FS for his data computation. So, the new profit in this scenario is (assuming allocation of all other patients remains the same):
    \begin{multline}
        \Delta_{t,2} = (P'+1)m + (P - P' - 1)l - kF \\ - g\sum_{p \in P}q_{p,t}\beta_{p,t} - g\beta_{p',t}.
    \end{multline}
    Now, $\Delta_{t,2} - \Delta_{t,1}$ is given by:
    \begin{equation}\label{difference}
        \Delta_{t,2} - \Delta_{t,1} = m - l - g\beta_{p',t}.
    \end{equation}
    From Eqs. (\ref{AdditionalConstraint}) and  (\ref{difference}), we can conclude that:
    \begin{equation}
        \Delta_{t,2} - \Delta_{t,1} \geq 0.
    \end{equation}
    As we increase the number of patients that utilize FSs for their computation, the profit also increases or remains constant. Hence, proved.
\end{proof}

\section{Problem Formulation}\label{Sec4}

In the proposed system, we are considering two factors: the profit of the medical center and the cost of patients. We know that a medical center will try to maximize its profit under the condition that no patient will face any delay in monitoring. Moreover, the patients would want to minimize their cost to ensure that they are being properly monitored. However, both objectives cannot be achieved at the same time. Thus, we are considering utility defined as the linear combination of profit of medical center and the cost of patients as follows:
\begin{equation}\label{UtilityBeyondI}
    U(t) = \lambda_{1}\Delta(t) - \lambda_{2}J(t),
\end{equation}
where, $\lambda_{1}$ and $\lambda_{2}$ are the weights assigned to the profit of the medical center and the cost of the patients respectively. The weights are taken as inverse units of the profit and the latency cost respectively, so that utility becomes unit-less. The weights are dependent on the system requirements and should be considered accordingly. That means, if the system is more profit aware then, $\lambda_1 > \lambda_2$, or if the system is more criticality aware then, $\lambda_1 < \lambda_2$, or if it is equally balancing between the two, then $\lambda_1 = \lambda_2$. Moreover, we are considering a constraint on the permissible latency defined as $\delta$ so as to ensure that no patient would face a delay of more than $\frac{\delta}{\rho_{p,t-1}^{c}}$. Furthermore, we have considered if the criticality of the patient is more, then the permissible latency is less. Thus, the optimization problem is formulated as follows:
\begin{equation}\label{ProblemBeyondI}
    \argmax_{\mathbb{H}_{t}}
    \hspace{1mm}U(t)
\end{equation}

Subject to the constraints:
\begin{equation}\label{cons1}
    \lambda_{1} \geq 0, \lambda_{2} \geq 0,
\end{equation}
\vspace{-10pt}
\begin{equation}\label{cons2}
    \lambda_{1} + \lambda_{2} = 2,
\end{equation}
\vspace{-10pt}
\begin{equation}\label{TimeBound}
    \forall p \in \mathbb{P}, \sum_{f'=1}^{F}h_{p,t}^{f'}T^{tr}_{p,f,t}+T^{c,f}_{p,t}(\mathbb{H}_{t}) +  u_{p,t} T^{c,l}_{p,t}\leq \frac{\delta}{\rho^{c}_{p,t-1}},
\end{equation}
\vspace{-10pt}
\begin{equation}\label{lmax}
    l \leq l_{max},
\end{equation}
\vspace{-10pt}
\begin{equation}\label{mmax}
    m \leq m_{max},
\end{equation}
\vspace{-10pt}
\begin{equation}\label{lmbound}
     m - l \geq g\beta_{p,t}^{max} + \frac{kF}{P},
\end{equation}
\vspace{-10pt}
\begin{equation}\label{cons3}
    \forall p \in \mathbb{P}, \sum_{f \in \mathbb{F}} h_{p,t}^{f} = q_{p,t}.
\end{equation}
The constraints given in Eqs. (\ref{cons1}) and (\ref{cons2}) are the bounds on the weights. Eq. (\ref{TimeBound}) is the latency constraint. Eqs. (\ref{lmax}) and (\ref{mmax}) put a constraint on the service charge. Eq. (\ref{lmbound}) refers to the additional constraint as defined in Eq. (\ref{AdditionalConstraint}). Eq. (\ref{cons3}) ensures that every patient is allocated at most one FS.

The formulated problem in Eqs. (\ref{ProblemBeyondI}-\ref{cons3}) is a Binary Integer Programming problem in $\mathbb{H}_{t}$ decision variables, that is generally NP-hard to solve as its feasibility problem is strongly NP-complete \cite{lewis1983computers}. Due to the high  conditionality and hardness of the  formulated  problem,  this  paper  proposes  a  complete  framework  to provide a sub-optimal solution for the maximization problem based on a swapping-based heuristic in the following section.

\section{Proposed Solution}\label{Sec6} 

To save the high computation charge at FS, each patient would like to compute the task at LD itself. However, while doing so, they may not meet the latency constraint, Eq. (\ref{TimeBound}). So, we have to prioritise these patients for utilizing the FSs. Thus, a sub-problem here is to allocate the FSs to such patients. Let $\mathbb{P}^{v}$ be the set of patients that violate the latency constraint if their data is computed at the LD. Formally, 
\begin{equation}
    \mathbb{P}^{v} = \{p \in \mathbb{P}: \rho^{c}_{p,t-1}T_{p,t}^{c,l} > \delta \}
\end{equation}

As per Lemma (\ref{Claim}), we can say that the profit only depends on patients who utilize the FSs and,  increasing the number of patients who utilize FSs increases the earned profit of medical center. So, if all the patients in $\mathbb{P}^{v}$ utilize FSs, the profit does not depend on how they are allocated the FSs. Thus, the utility depends only on the cost function as defined in Eq. (\ref{PatientCost}). 

In this sub-problem, we offload the patient's data to one FS. After allocation of patients in the set $\mathbb{P}^{v}$ (patients violating latency constraint), our next objective is to allocate a subset of the remaining patients such that it maximizes the utility under the system constraints. We can notice that it is not possible to allocate all the patients to FSs due to the limited resources. Doing so may result in violation of the latency constraint Eq. (\ref{TimeBound}) and could increase the patients' cost drastically, which would result in low utility. Consider the constraint given in Eq. (\ref{TimeBound}), and let $n_{p,f,t}^{max}$ be the maximum number of patients (see Theorem \ref{NumberConstraintTheorem}) that can utilize the FS $f$ for their computation if patient $p$ utilizes it. 

\begin{theorem}\label{NumberConstraintTheorem}
The maximum number of patients that can utilize the FS $f$ for their computation, if patient $p$ utilizes that FS, can be given as:
\begin{equation}\label{NumberConstraint}
    n_{p,f,t}^{max} = \Bigl\lfloor \left(\frac{\Gamma}{\beta_{p,t}}\right) \left(\frac{\delta}{\rho_{p,t-1}^{c}} - \frac{\eta_{p,t}}{\Omega \log_{2}\left( 1 + \frac{W^{bey}_{p,f,t}}{N^{bey}_{t}} \right)}\right) \Bigr\rfloor
\end{equation}
\end{theorem}

\begin{proof}
According to Eq. (\ref{TimeBound}), if a patient $p$ utilizes FS $f$, then,
\begin{equation}
    T^{tr}_{p,f,t} + T^{c,f}_{p,t}(\mathbb{H}_t) \leq \frac{\delta}{\rho_{p,t-1}^{c}},
\end{equation}
Putting values of $T^{tr}_{p,f,t}$ and $T^{c,f}_{p,t}(\mathbb{H}_t)$ from Eqs. (\ref{TimeBeyond}) and (\ref{TimeFog}) respectively, we get
\begin{equation}
    \frac{\eta_{p,t}}{BR_{p,f,t}} + \frac{\beta_{p,t}}{\gamma_{p}(\mathbb{H}_{t})} \leq  \frac{\delta}{\rho_{p,t-1}^{c}},
\end{equation}
where, $\gamma_{p}(\mathbb{H}_{t}) = \frac{\Gamma}{{n'}_{p,f,t}}$, where ${n'}_{p,f,t}$ are the number of patients utilizing FS $f$ including $p$. After solving the inequality and putting the value of $BR_{p,f,t}$, we get,

\begin{equation}
    {n'}_{p,f,t} \leq \left(\frac{\Gamma}{\beta_{p,t}}\right) \left(\frac{\delta}{\rho_{p,t-1}^{c}} - \frac{\eta_{p,t}}{\Omega \log_{2}\left( 1 + \frac{W^{bey}_{p,f,t}}{N^{bey}_{t}} \right)}\right),
\end{equation}
where, ${n'}_{p,f,t}$ is an integer. So, the maximum value of ${n'}_{p,f,t}$ is the greatest integer value of the right-hand side expression. Hence, proved.

\end{proof}

To solve the formulated problem, we have proposed Utility Maximization Patient Monitoring (UMPM) algorithm. The main idea behind the proposed heuristic is, to begin with, an initial allocation and then, re-positioning the patients by swapping their positions in order to achieve higher utility, iteratively, as shown if Fig. \ref{Flow_chart}. The elaboration of each sub-algorithm is described as following: 

\begin{figure}[h!]
	\centering
	\vspace{0pt}
	\includegraphics[height=7cm, width=8cm]{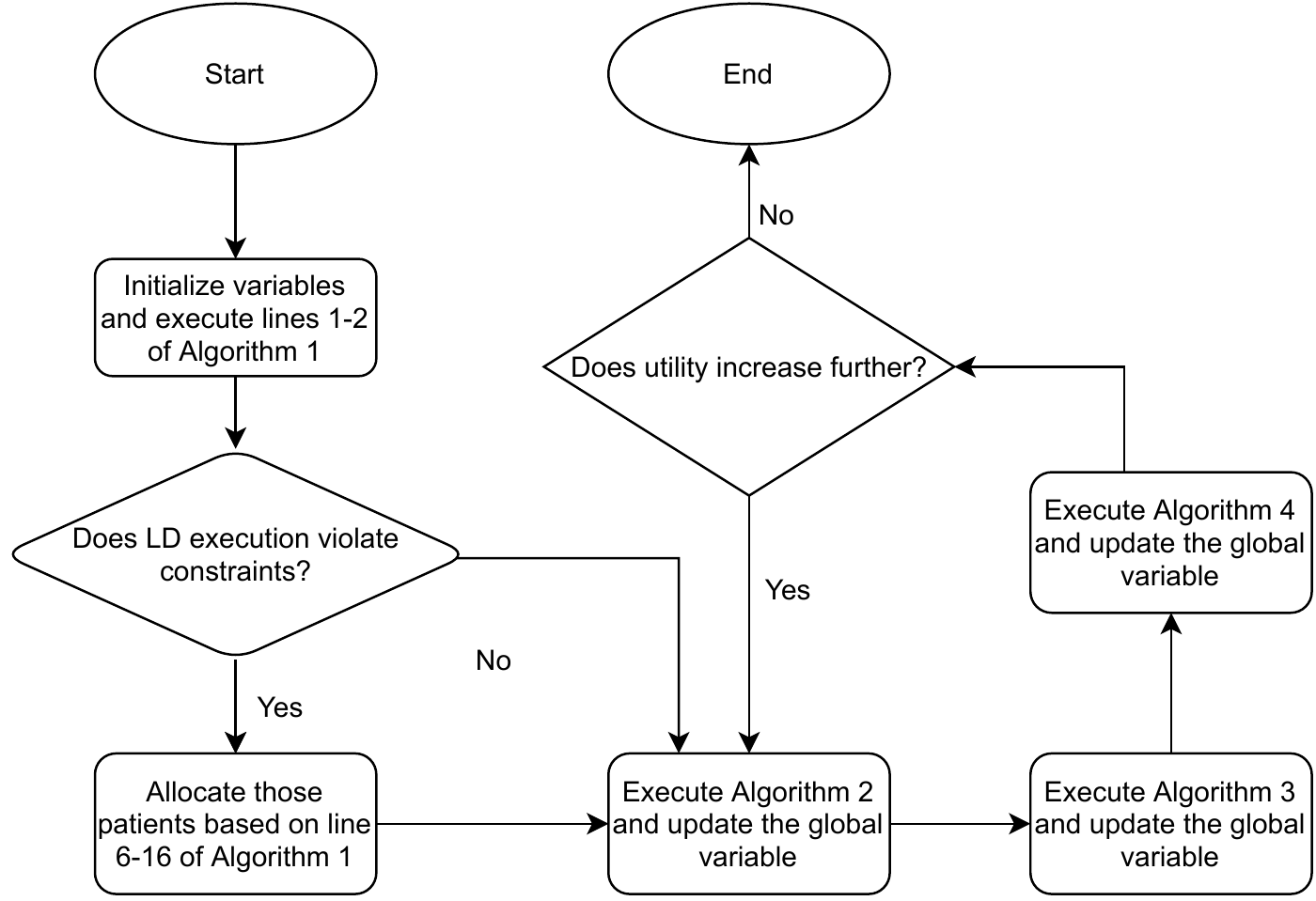}
	\caption{Flow chart of the proposed heuristic.}
	\label{Flow_chart}
	\vspace{0pt}
\end{figure}

\subsection{UMPM Algorithm}

Keeping the allocation of other patients as it is, we define $U_{diff}$\footnote{The profit is changed only for the patient $p$ and the patient $p$ affects the cost function with a value of the difference of $p$'s local computation time and its transmission and fog computation time. The computation time of the remaining patients using $f$ also changes.} as the difference between utility before and after a patient $p$ is allocated an FS $f$ as follows:
\begin{multline}\label{Udiff}
    U_{diff} = \lambda_{1}(m - l - g\beta_{p,t}) +  \\ \lambda_{2} \left(\rho^{c}_{p,t-1}\left(\frac{\beta_{p,t}}{\Upsilon} - \left( \frac{\beta_{p,t}(n_{f}+1)}{\Gamma} + \frac{\eta_{p,t}}{BR_{p,f,t}} \right)\right)\right) \\
    - \lambda_{2}\sum_{p' \in  P^{f}}\frac{\rho_{p',t-1}^{c}\beta_{p',t}}{\Gamma}).
\end{multline}
The set of patients that utlize FS $f$ can be calculated as:
\begin{equation}
    \mathbb{P}^{f} = \{p \in \mathbb{P}: q_{p,t} = 1\}.
\end{equation}
Moreover, the number of patients utilizing FS $f$ can be estimated as:
\begin{equation}
    n^{f} = \sum_{p = 1}^{P} h_{p,t}^{f}.
\end{equation}

\begin{algorithm}[h!]
\caption{ UMPM Algorithm}
\KwIn{ $\Gamma$, $\Upsilon$, $g,m,l,\delta$, $\mathbb{H}_{t} = \{0\}$, $\Omega$, $N_{t}^{bey}$; $\forall p \in \mathbb{P}$: $\rho_{p,t-1}^{c}$, $\beta_{p,t}$, $\eta_{p,t}$; $\forall p \in \mathbb{P}, \forall f \in \mathbb{F}$: $W_{p,f,t}^{bey}$; $\forall f \in \mathbb{F}: \mathbb{P}^{f}, n^{f} = 0$; $temp_p,temp_f,U^{max}_{diff} = -\infty$}
\KwOut{Allocation Strategy ($\mathbb{H}_{t}$).}
\label{Main}
Calculate $n_{p,f,t}^{max}$ for all the patients and FS using Theorem \ref{NumberConstraintTheorem}\;
 Calculate local computation time for the patients using Eq. (\ref{CompTimeLocal})\;
 \For{$p\gets1$ \KwTo $P$}{
    \If{$T^{c,l}_{p,t} > \frac{\delta}{\rho_{p,t-1}^{c}}$}{
        insert $p$ into $\mathbb{P}^{v}$
    }
 }
 Sort patients in set $\mathbb{P}^{v}$ in the order of decreasing criticalities.\;
 \For{$p \in \mathbb{P}^{v}$}{
    \For{$f \in F$}{
         \If{$n_{f} \geq \min_{p' \in \mathbb{P}^{f} \bigcup \{p\}}n_{p',f,t}^{max}$}{
                $\Continue$
            }
        Calculate $U_{diff}$ as per Eq. (\ref{Udiff})\;
        \If{$U_{diff} > U_{diff}^{max}$}{
            $U_{diff} \gets U_{diff}^{max}$\;
            $temp_p \gets p$\;
            $temp_f \gets f$\;
        }        
    }
    Allocate $temp_p$ to $temp_f$ and update variables accordingly;
 }
 Run Algorithm \ref{Swapping2}\;
 Run Algorithm \ref{Swapping}\;
  $\mathbb{P}^{rem} \gets \mathbb{P} - \mathbb{P}^{v}$   \tcp*{Update $\mathbb{P}^{rem}$ as the set of patients who have not yet assigned an FS}
  Run Algorithm \ref{Greedy}\;
 \Repeat{Utility does not increase}{
    Run Algorithm \ref{Swapping2}\;
    Run Algorithm \ref{Swapping}\;
    Run Algorithm \ref{Greedy}\;
    Update $\mathbb{P}^{rem}$\;
 }
\end{algorithm}
\setlength{\textfloatsep}{0pt}
The UMPM algorithm begins by calculating the constraint parameter $n_{p,f,t}^{max}$, as defined in Eq. (\ref{NumberConstraint}). Then, it selects a subset of patients that violates the latency constraint if their data is computed on LD. Further, the algorithm sorts the patients in $\mathbb{P}^{v}$ in the order of their decreasing criticalities. By doing so, the patients with higher criticalities are given priority by the algorithm, and thus, the algorithm will be able to accommodate all the patients in $P^{v}$ on FS. After that, re-allocation is done using Algorithms \ref{Swapping2} and \ref{Swapping}. After the initial allocation, the algorithm constructs the set $\mathbb{P}^{rem}$, which is the set of patients who have not been allocated any FS. It calls Algorithm \ref{Greedy} to allocate more patients on FS. It then calls Algorithms \ref{Swapping2} and \ref{Swapping} which re-position the patients on the FSs. After that, Algorithm \ref{Greedy} is further called to allocate more patients on FS so as to increase utility. This process is repeated until there is no such possibility of increment in the utility (Fig. \ref{Flow_chart}). Moreover, we have considered $\mathbb{H}_t$ as a global matrix available at cloud server and accessed by all the proposed Algorithms \ref{Main}-\ref{Greedy}. 

The reason for having Algorithms \ref{Swapping2} and \ref{Swapping} is to obtain better utility by swapping the association between FSs and patients as described below: 

\subsection{Two Way Swap based Algorithm}
The utility difference due to two way swap, $J_{diff}^{tr}$\footnote{When two patients $p$ and $p'$ utilizing servers $f$ and $f'$ respectively are swapped, then the change in utility is caused by the difference of their transmission latencies and their computation latencies, as considered in Eq. (\ref{JdiffTr}).} can be calculated as: 
\begin{multline}\label{JdiffTr}
    J_{diff}^{tr} = \frac{\rho_{p,t-1}^{c}\eta_{p,t}}{BR_{p,f,t}} - \frac{\rho_{p,t-1}^{c}\eta_{p,t}}{BR_{p,f',t}} +
    \\ \frac{\rho_{p',t-1}^{c}\eta_{p',t}}{BR_{p',f',t}} - \frac{\rho_{p',t-1}^{c}\eta_{p',t}}{BR_{p',f,t}} + \\ \frac{\rho_{p,t-1}^{c}\beta_{p,t}n_{f}}{\Gamma} - \frac{\rho_{p,t-1}^{c}\beta_{p,t}n_{f'}}{\Gamma} +
    \\ \frac{\rho_{p',t-1}^{c}\beta_{p',t}n_{f'}}{\Gamma} -
    \frac{\rho_{p',t-1}^{c}\beta_{p',t}n_{f}}{\Gamma}.
\end{multline}

\begin{algorithm}
\caption{Two way swap}
\label{Swapping2}
\KwIn{Globally accessible $\mathbb{H}_{t}$ , Information of all patients (as per Algorithm \ref{Main}) and FSs.}
\KwOut{$\mathbb{H}_{t}$}
\Repeat{No swap increases utility}{
    \For{$f\gets1$ \KwTo $F$}{
        \For {every $p \in  \mathbb{P}^{f}$}{
            \For{${f'}\gets 1$ \KwTo $F$}{
                \If{$f' == f$}{
                    \Continue
                }
                \For{every $p' \in \mathbb{P}^{f'}$}{
                    \If{$n_{f'} > n_{p,f',t}^{max}$ or $n_{f} > n_{p',f,t}^{max}$}{
                        \Continue
                    }
                    \If{$J_{diff}^{tr} > 0$}{
                        Swap $p$ and $p'$\;
                        Update all the values accordingly\;
                        Go to the Repeat loop\;
                    }
                }
               }
                   }
            }
}
\end{algorithm}

%explanation
 The algorithm at each iteration picks a pair of patients already allocated to different FSs (lines 2-7). It then checks whether swapping the position of these two patients can increase utility or not (line 10). If the utility can be increased by satisfying the constraints (line 8), the patients are swapped (lines 11-12). It repeats the process until there is no such pair of patients (line 13). The convergence proof can be found in Section \ref{analysis}.

%--------------------------------------------------

\subsection{One Way Swap based Algorithm}
The utility difference due to one way swap, $J_{diff}$\footnote{When a patient $p$ utilizing FS $f$ is allocated FS $f'$, we can observe that profit does not change. The change in the cost function can be calculated as the difference between the transmission times when $p$ uses $f$ and $f'$. Also, the computation time of the patient $p$ changes. Other than that, the computation time of the patients utilizing $f$ and $f'$, other than $p$ changes. All these changes are considered in $J_{diff}$.} is calculated as follows:
\begin{multline}\label{Jdiff}
    J_{diff} = \frac{\rho_{p,t-1}^{c}\eta_{p,t}}{BR_{p,f,t}} - \frac{\rho_{p,t-1}^{c}\eta_{p,t}}{BR_{p,f',t}} 
    + \frac{\rho_{p,t-1}^{c}\beta_{p,t}(n_f - n_{f'} -  1)}{\Gamma} \\ + \sum_{p' \in \mathbb{P}^{f} \backslash \{p\}}\frac{\rho_{p',t-1}^{c}\beta_{p',t}}{\Gamma} - \sum_{p' \in \mathbb{P}^{f'}}\frac{\rho_{p',t-1}^{c}\beta_{p',t}}{\Gamma}.
\end{multline}

\begin{algorithm}
\caption{One Way Swap}
\label{Swapping}
\KwIn{Globally accessible $\mathbb{H}_{t}$, Information of all patients (as in Algorithm \ref{Main}) and FS.}
\KwOut{$\mathbb{H}_{t}$}
\Repeat{No swap increases utility}{
    \For{$f\gets1$ \KwTo $F$}{
        \For {every $p \in  \mathbb{P}^{f}$}{
            \For{${f'} \gets 1$ \KwTo $F$}{
                \If{$f' == f$}{
                    \Continue
                }
                Compute $J^{diff}$ according to Eq. (\ref{Jdiff})\;
                \If{$J_{diff} > 0$ and $n_{f'} + 1 \leq \min_{p' \in \mathbb{P}^{f'} \bigcup \{p\}}(n_{p',f',t}^{max})$}{
                    Add $p$ to $\mathbb{P}^{f'}$ and remove $p$ from $\mathbb{P}^{f}$\;
                    Update the values correspondingly\;
                    Go to the Repeat loop\;
                }
            }
           
        }
        
    }
}
\end{algorithm}

 The algorithm at each iteration picks a patient allocated to an FS and checks if assigning that patient a different FS can increase utility or not, satisfying the constraints (lines 2-8). If the utility can be increased by satisfying the constraints, the patients are assigned to a different FS (lines 9-10). Similar to the two-way swapping, this algorithm also repeats the process until there is no such pair of patients (line 11). The convergence proof can be found in Section \ref{analysis}.

%------------------------------------------------------

%start here

\subsection{Patient-FS Allocation Algorithm}
 The algorithm considers all patient-FS pairs at each iteration and selects the one that increases the utility by maximum value. In this way, the algorithm selects a subset of the patients given as input and allocates them to the FSs satisfying the constraints and maximising the utility value. Algorithm terminates when there is no improvement in the utility value compared to utility obtained in previous iteration. The following Lemma \ref{LPSAA} establishes an iterative utility correlation across different iteration of Algorithm \ref{Greedy}.
\begin{lemma}\label{LPSAA}
    Let $U_{diff,i}^{max}$ be the $U_{diff}^{max}$ calculated by the algorithm at $i^{th}$ iteration, then $U_{diff,i}^{max} \geq U_{diff,i+1}^{max}$. In other words, the maximum utility difference decreases with each iteration of Algorithm \ref{Greedy}.
\end{lemma}

\begin{proof}
Let $p$ be the patient assigned to FS $f$ at $i^{th}$ iteration and $p'$ be the patient assigned to FS $f'$ at $({i+1})^{th}$ iteration. Then, consider two following cases:
    
    \textit{Case 1: $f = f'$}
    \begin{multline}\label{lemma2eq1}
    U_{diff,i}^{max} = \lambda_{1}(m - l - g\beta_{p,t})  + \\ \lambda_{2} \left( \rho^{c}_{p,t-1}\left(\frac{\beta_{p,t}}{\Upsilon} - \left( \frac{\beta_{p,t}(n_{f}+1)}{\Gamma} + \frac{\eta_{p,t}}{BR_{p,f,t}} \right)\right) \right)\\
    - \lambda_{2} \sum_{p'' \in  P^{f}}\frac{\rho_{p'',t-1}^{c}\beta_{p'',t}}{\Gamma}.
    \end{multline}
    Also,
    \begin{multline}\label{lemma2eq2}
    U_{diff,i+1}^{max} = \lambda_{1}(m - l - g\beta_{p',t})  + \\ \lambda_{2} \left(\rho^{c}_{p',t-1}\left(\frac{\beta_{p',t}}{\Upsilon} - \left( \frac{\beta_{p',t}(n_{f}+2)}{\Gamma} + \frac{\eta_{p',t}}{BR_{p',f,t}} \right)\right) \right) \\
    - \lambda_{2}\sum_{p'' \in  P^{f} \bigcup {p}}\frac{\rho_{p'',t-1}^{c}\beta_{p'',t}}{\Gamma},
    \end{multline}
    where $n^{f}$ are the number of patients utilizing $f$ before $i^{th}$ iteration and similarly $\mathbb{P}^{f}$ is the set of such patients. On subtracting Eq. (\ref{lemma2eq1}) from Eq. (\ref{lemma2eq2}), it is clear that,
    \begin{equation}
    U_{diff,i+1}^{max} \le U_{diff,i}^{max}. 
    \end{equation}
    
    \textit{Case 2: $f \neq f'$}
    
    In this case, as both the FSs are different, if $U_{diff,i+1}^{max}$ would have been greater than $U_{diff,i}^{max}$, then the algorithm would have picked $p'$ at the $({i+1})^{th}$ iteration only, but that is not the case. Hence $U_{diff,i+1}^{max} \leq U_{diff,i}^{max}$. 
    
    Combining both the cases, we conclude,
    $U_{diff,i+1}^{max} \leq U_{diff,i}^{max}$. Hence, proved.
\end{proof}
\begin{algorithm}
\caption{Patient-FS Allocation}
\label{Greedy}
\KwIn{Globally accessible $\mathbb{H}_{t}$, Set of Patients, $P^{rem}$, $flag = 0$, $temp_p$, $temp_f$, $U^{max}_{diff} = 0$}
\KwOut{Allocation Strategy ($\mathbb{H}_{t}$).}
 \RepTimes{$|\mathbb{P}^{rem}|$}{
    \For{every $p \in \mathbb{P}^{rem}$}{
        \If{$q_{p,t}$ == 1}{
            \Continue\;
        }
        \For{$f\gets1$ \KwTo $F$}{
            \If{$n_{f} \geq \min_{p' \in \mathbb{P}^{f} \bigcup \{p\}}n_{p',f,t}^{max}$}{
                $\Continue$
            }
            Calculate $U_{diff}$ as per Eq. (\ref{Udiff})\;
            \If{$U_{diff} > U_{diff}^{max}$}{
                $flag \gets 1$\;
                $U_{diff} \gets U_{diff}^{max}$\;
                $temp_p \gets p$\;
                $temp_f \gets f$\;
            }
        }
    }
    \If{$flag == 0$}{
        $\Break$
    }
    Assign patient $temp_p$ to FS $temp_f$\;
    Update $n^{temp_f}$ and $\mathbb{P}^{temp_f}$\;
    Update $\mathbb{H}_{t}$\;
 }

\end{algorithm}

\subsection{Illustration of UMPMA}
Let there be 3 FSs and 10 patients. We illustrate the proposed heuristic using a randomly generated example under the simulation parameters as mentioned in Table \ref{sim_param}. The yellow and blue colours represent patients and FSs respectively, in subsequent figures.

\begin{figure}[h!]
	\centering
	\includegraphics[height=4cm, width=8cm]{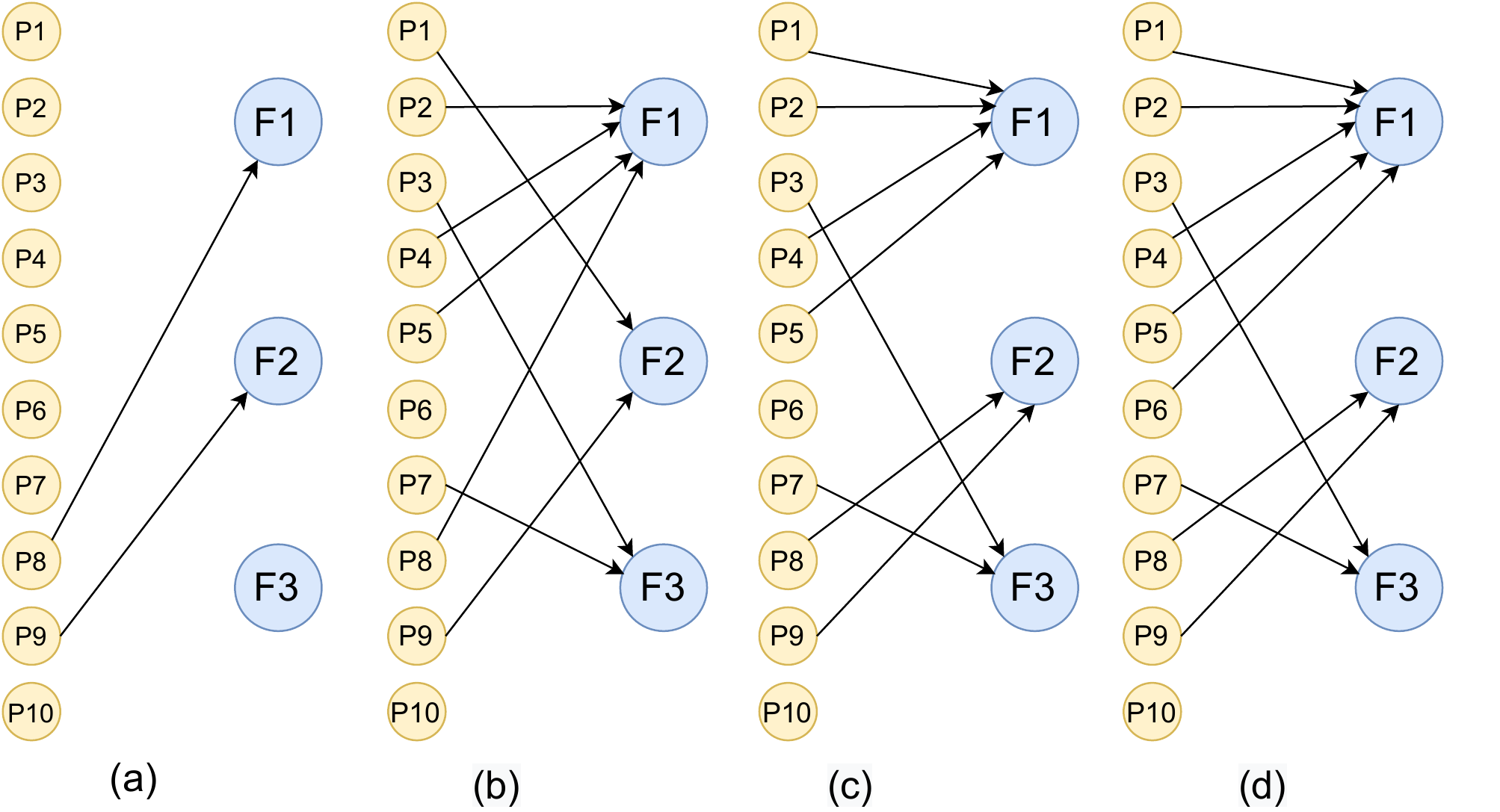}
	\caption{(a) Initial allocation. (b) Updated allocation after execution of Algorithm \ref{Greedy}. (c) Updated allocation in next iteration. (d) Final allocation.}
	\label{illustration_phase_1}
% 	\vspace{0pt}
\end{figure}

Let Fig. \ref{illustration_phase_1} (a) shows an allocation of patients to FSs after execution of lines 1-18 (including Algorithms \ref{Swapping2}-\ref{Swapping}). We can see that P8 and P9 are allocated to F1 and F2 respectively. 
Moreover, Algorithm \ref{Greedy} executes to see any possible update in the allocation for improving utility value. In this algorithm, an allocation is found out while maximizing overall utility value. Thus, we get an outcome as shown in Fig. \ref{illustration_phase_1} (b). Next, after re-iterating the given swapping algorithms, Fig. \ref{illustration_phase_1} (c) is obtained as an outcome improving the utility over the allocation given in Fig. \ref{illustration_phase_1} (b). Furthermore, after convergence of the proposed heuristic, we achieve Fig. \ref{illustration_phase_1} (d) as a final outcome maximizing the overall utility over all previous allocations. 

\section{Analysis of Proposed Heuristic}\label{analysis}
This section discusses convergence and time complexity of the proposed heuristic as follows.
\begin{lemma}\label{conv}
    The proposed UMPMA converges.
\end{lemma}

\begin{proof}
    The convergence of the UMPMA relies on the convergence of one-way and two-way swap algorithms. Both the swapping algorithms swaps only if utility increases and if there is no such swap possible, their execution is terminated. As the total possible combinations (patient-FS) are finite, hence the utility will also be a finite value. Thus, both the swapping mechanisms converge. Algorithm \ref{Greedy} converges because the number of iterations are bounded by a finite value i.e., the number of remaining patients. Further, the proposed heuristic repeatedly executes two-way swap, one-way swap and Patient-FS allocation schemes. Every iteration converges and the algorithm goes to the next iteration only when utility increases. A similar argument regarding the finiteness of the utility concludes the proof.
\end{proof}

\begin{theorem}\label{timecompelxity}
    The time complexity of UMPMA is $O(P^{2}F)$.
\end{theorem}

\begin{proof}
    The time complexity of the UMPMA depends on the complexity of the three sub-algorithms it calls. The time complexity of Algorithm \ref{Main} from lines (1-16) is $O(PF)$. Through amortized analysis, we can see that Algorithm \ref{Swapping2} considers $O(P^{2})$ pairs of patients. The algorithm repeats until it converges. Thus, the required number of iterations is bounded by a finite value. Thus, the time complexity of Algorithm \ref{Swapping2} is $O(P^{2})$. Similarly, through amortized analysis, the time complexity of Algorithm \ref{Swapping} is $O(PF)$ as it considers $O(PF)$ number of possibilities of swapping. In Algorithm \ref{Greedy}, the number of iterations are bounded by $O(P)$. In every iteration, $O(PF)$ pairs are considered. Thus, the time complexity of Algorithm \ref{Greedy} is $O(P^{2}F)$. Hence, the time complexity of proposed UMPMA is $O(max\{P^{2}F,P^{2},PF\})$, i.e., $O(P^{2}F)$.
\end{proof}

Performance study of the proposed heuristic is given in the following section over different simulation parameters.

\section{Performance Study}\label{Sec8}
The simulation setup and the parameters are given in the following:

\begin{table}[h]
\caption{Simulation Parameters}
\begin{tabular}{ |p{5cm}||p{2cm}|  }
 \hline
%  \multicolumn{2}{|c|}{Simulation Parameters} \\
%  \hline
 Parameter & Value\\
 \hline
 Number of Patients ($P$)  & 20-1000  \\
 Number of FS ($F$) & 2-200  \\
 Data Size ($\eta_{p,t}$) & [1,3] MB \\
 CPU Cycles ($\beta_{p,t}$) & [100,1000] Megacycles \\
 Time Constraint ($\delta$) & 250 ms \\
 Patient Criticality ($\rho^{c}_{p,t-1}$) & [0,1] \\
 Local Computation Price ($l$) & 100 units \\
 Fog Computation Price ($m$) & 200 units \\
 FS Charge per CPU Cycles ($g$) & 0.1 units \\
 Fixed Charge per FS ($k$) & 0 units \\
 Distance between LD and FS & [50,100] m \\
 Path loss factor & 3 \\
 Channel Bandwidth ($\Omega$) & 5 MHz \\
 Noise ($N^{bey}_{t}$) & -100 dBm \\
 Transmission power ($P^{bey}_{p,t}$) & 0.1 W \\
 Fog Computation Capacity ($\Gamma$) & 22.4 GHz \\
 Local Computation Capacity ($\Upsilon$) & 2.4 GHz \\
 Patients' Weight ($\lambda_1$) & 1 \\ 
 Medical Center's Weight ($\lambda_2$) & 1 \\
 \hline
\end{tabular}\label{sim_param}
\end{table}

\subsection{Simulation Setup}

\begin{figure*}[ht!]
 	\centering
 	\begin{subfigure}[b]{0.325\textwidth}
 		\centering
 		\includegraphics[width=\textwidth]{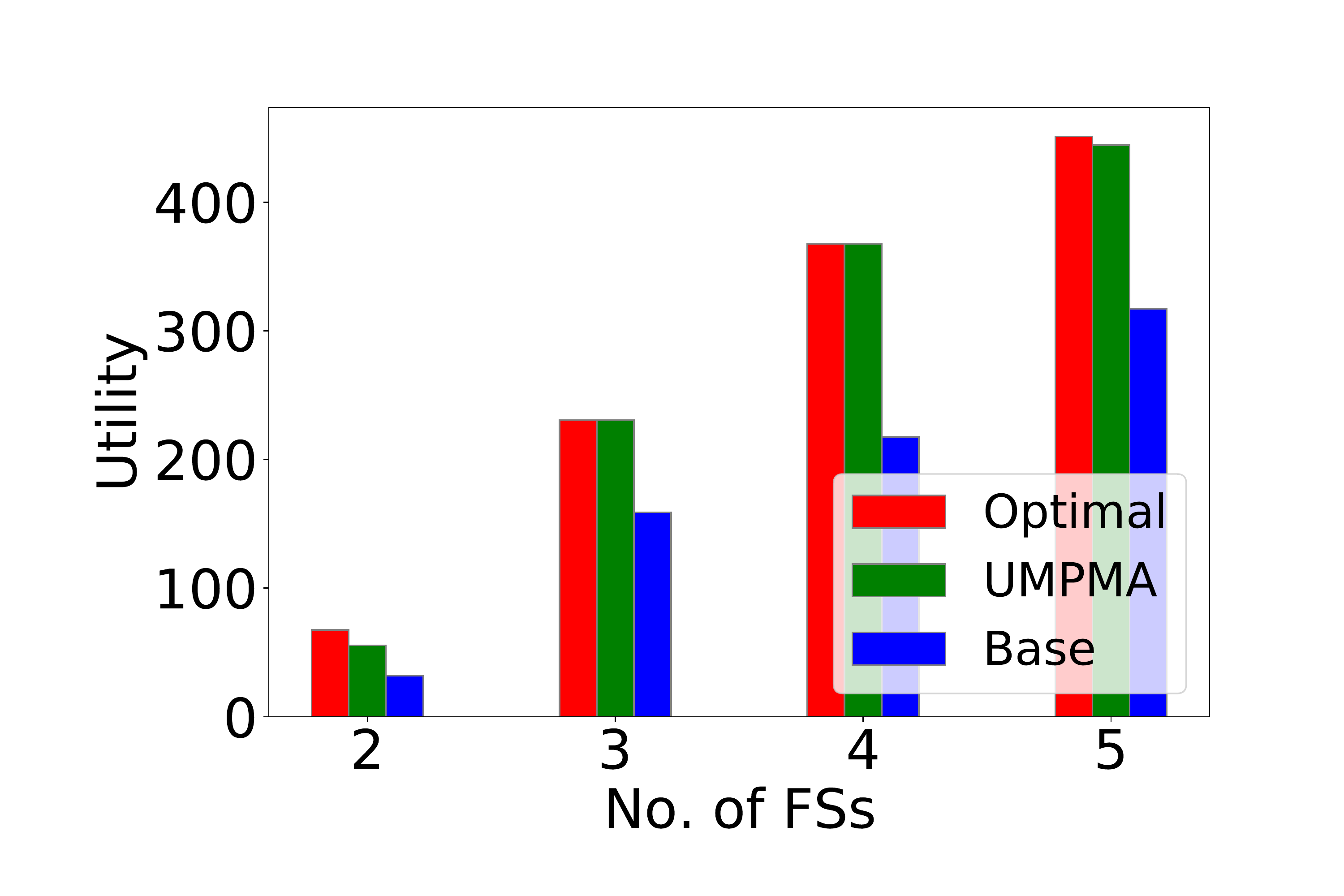}
 		\caption{No. of patients = 20.}
 		\label{plot_utility_20}
 	\end{subfigure}
 	\begin{subfigure}[b]{0.325\textwidth}
 		\centering
 		\includegraphics[width=\textwidth]{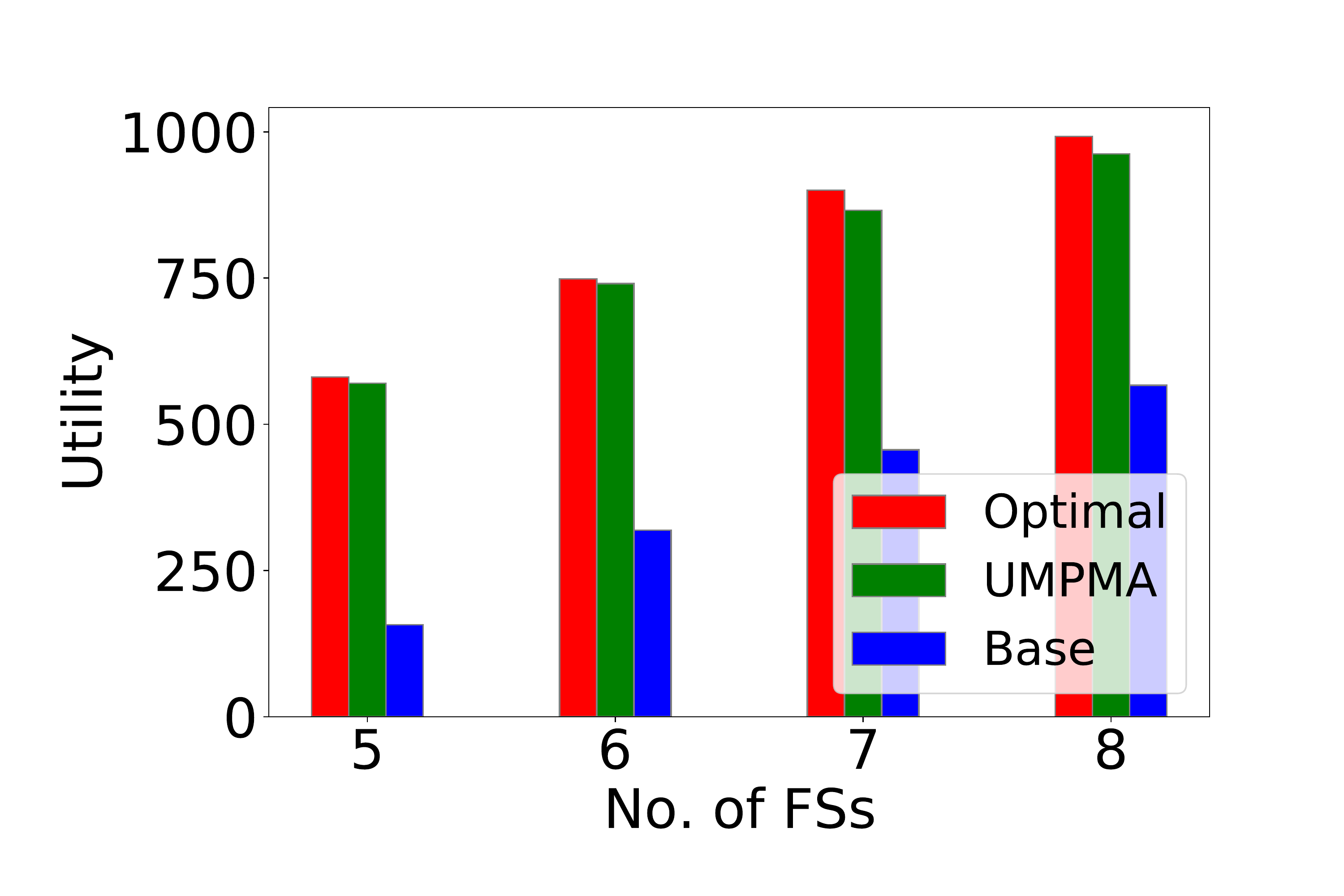}
 		\caption{No. of patients = 40.}
 		\label{plot_utility_40}
 	\end{subfigure}
 	\begin{subfigure}[b]{0.325\textwidth}
 		\centering
 		\includegraphics[width=\textwidth]{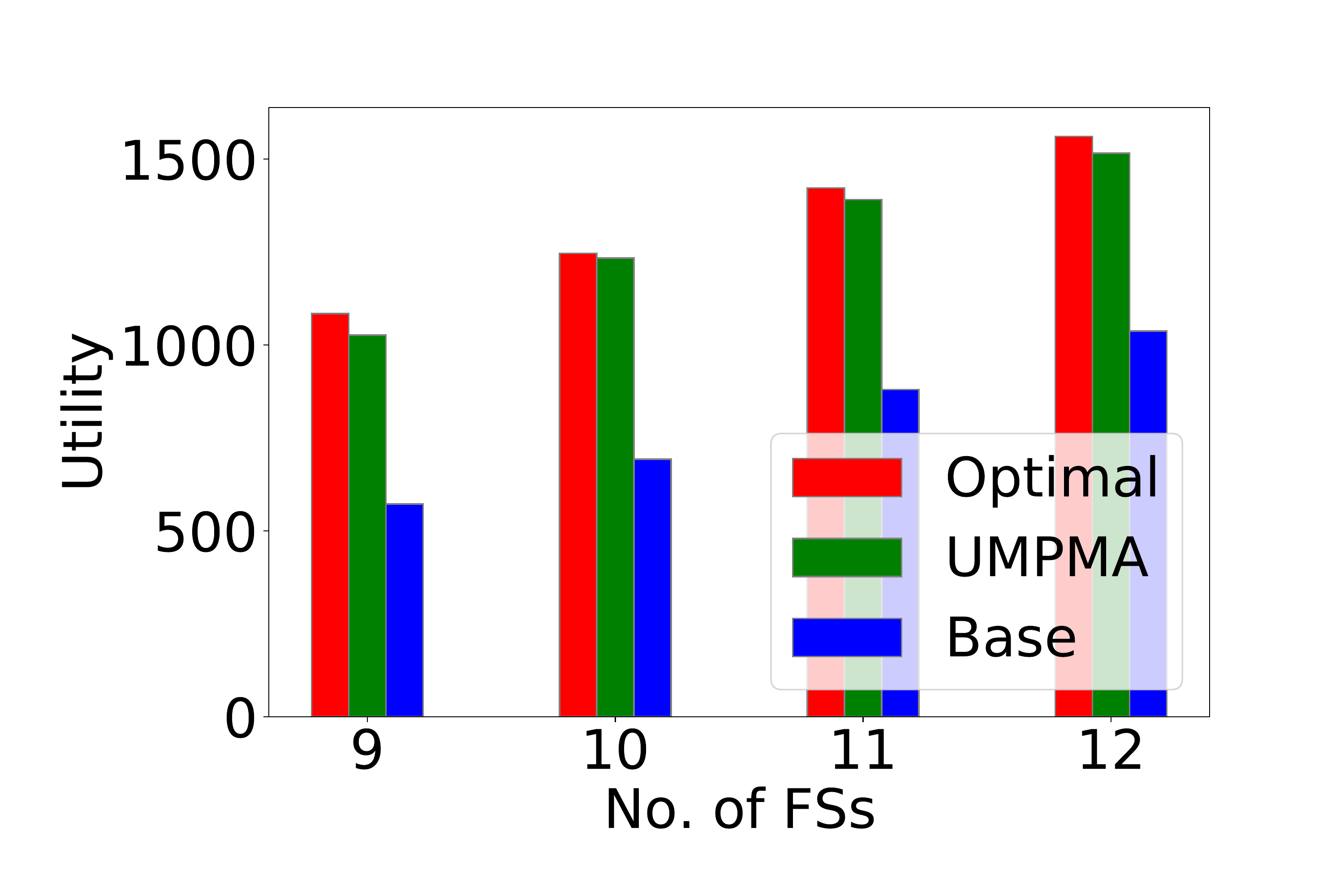}
 		\caption{No. of patients = 60.}
 		\label{plot_utility_60}
 	\end{subfigure}
 	\caption{Utility comparison among different schemes.}
 	\label{plot_utility}
 		\vspace{-0.1in}
 \end{figure*}
 
 \begin{figure*}[ht!]
 	\centering
 	\begin{subfigure}[b]{0.325\textwidth}
 		\centering
 		\includegraphics[width=\textwidth]{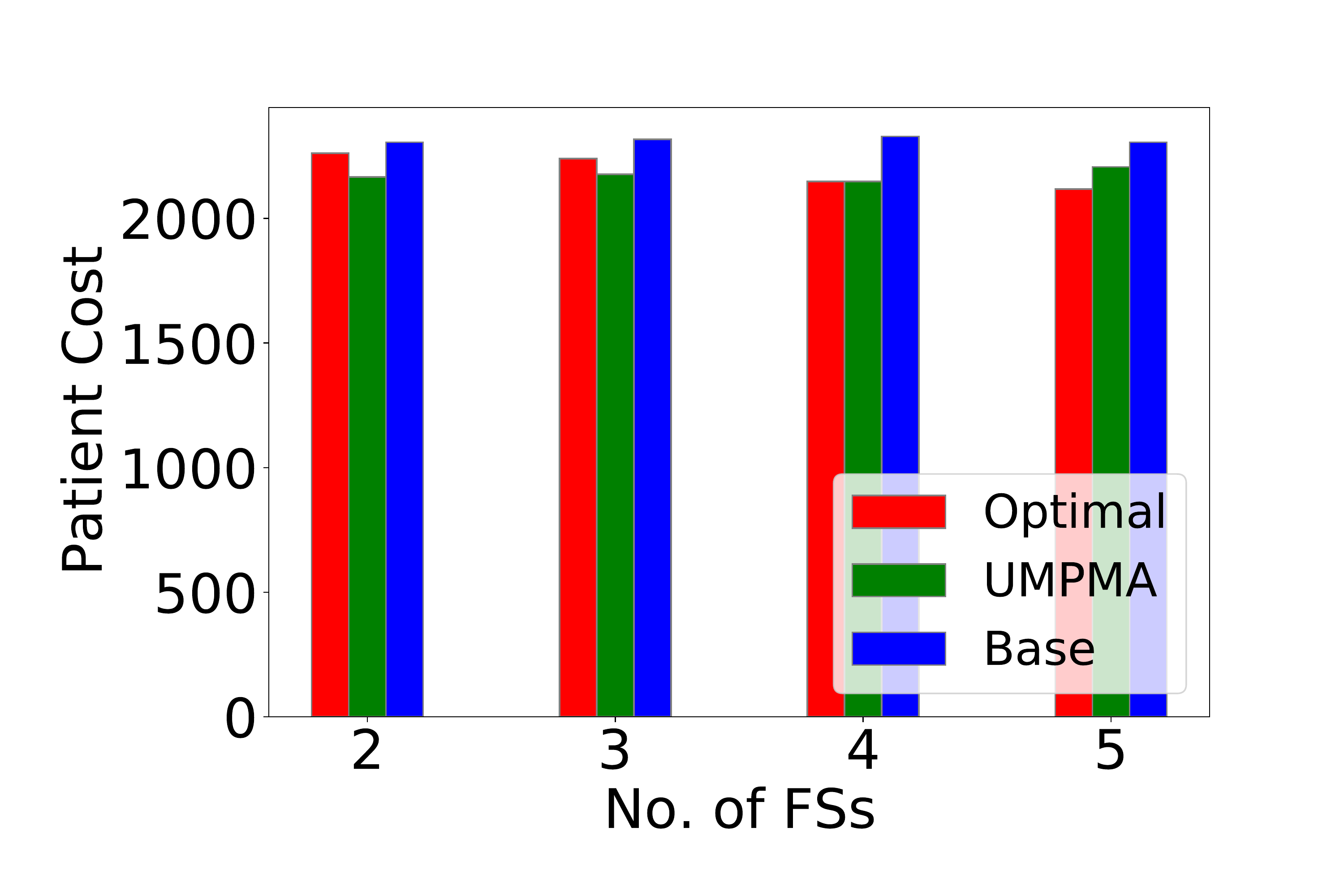}
 		\caption{No. of patients = 20.}
 		\label{plot_cost_20}
 	\end{subfigure}
 	\begin{subfigure}[b]{0.325\textwidth}
 		\centering
 		\includegraphics[width=\textwidth]{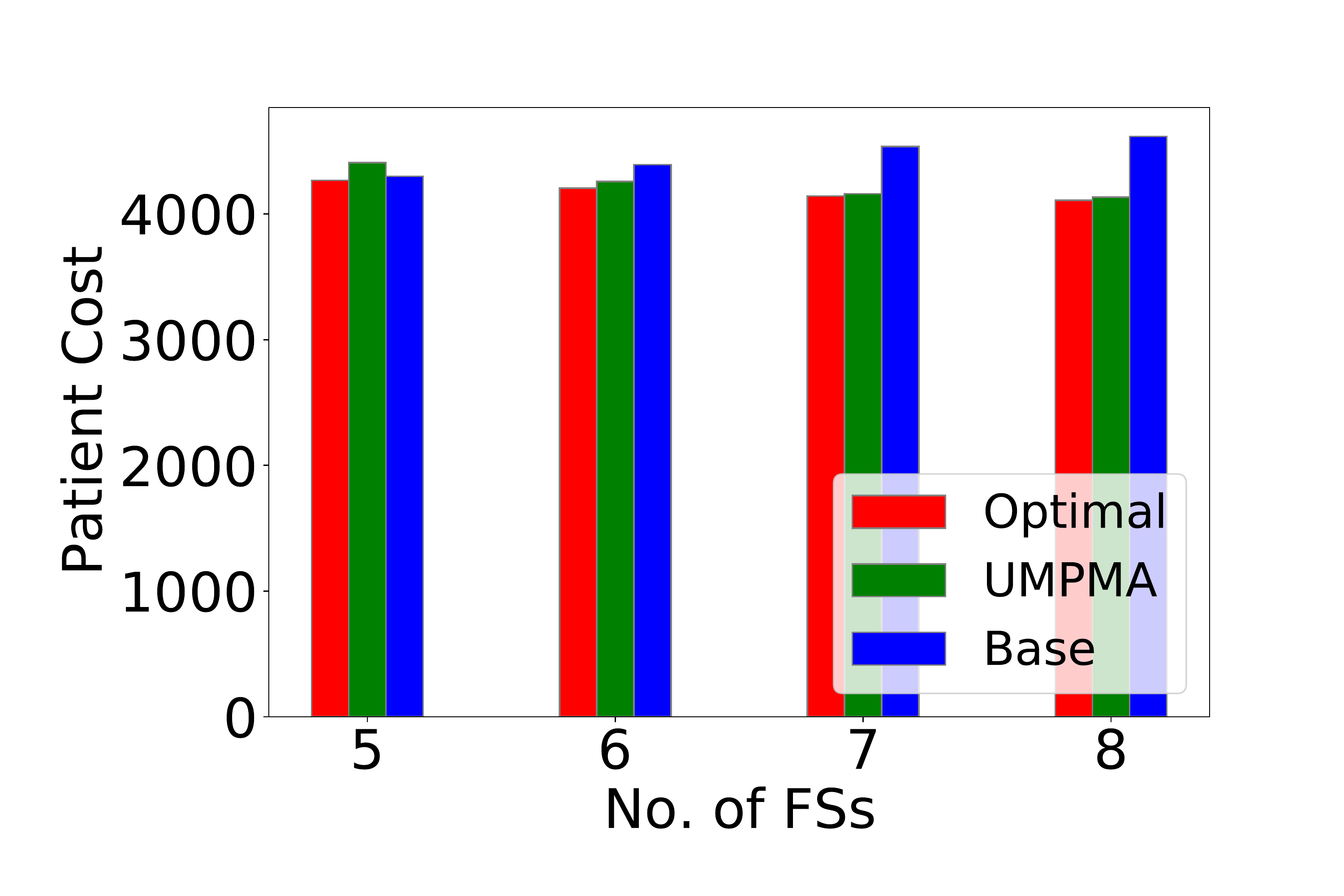}
 		\caption{No. of patients = 40.}
 		\label{plot_cost_40}
 	\end{subfigure}
 	\begin{subfigure}[b]{0.325\textwidth}
 		\centering
 		\includegraphics[width=\textwidth]{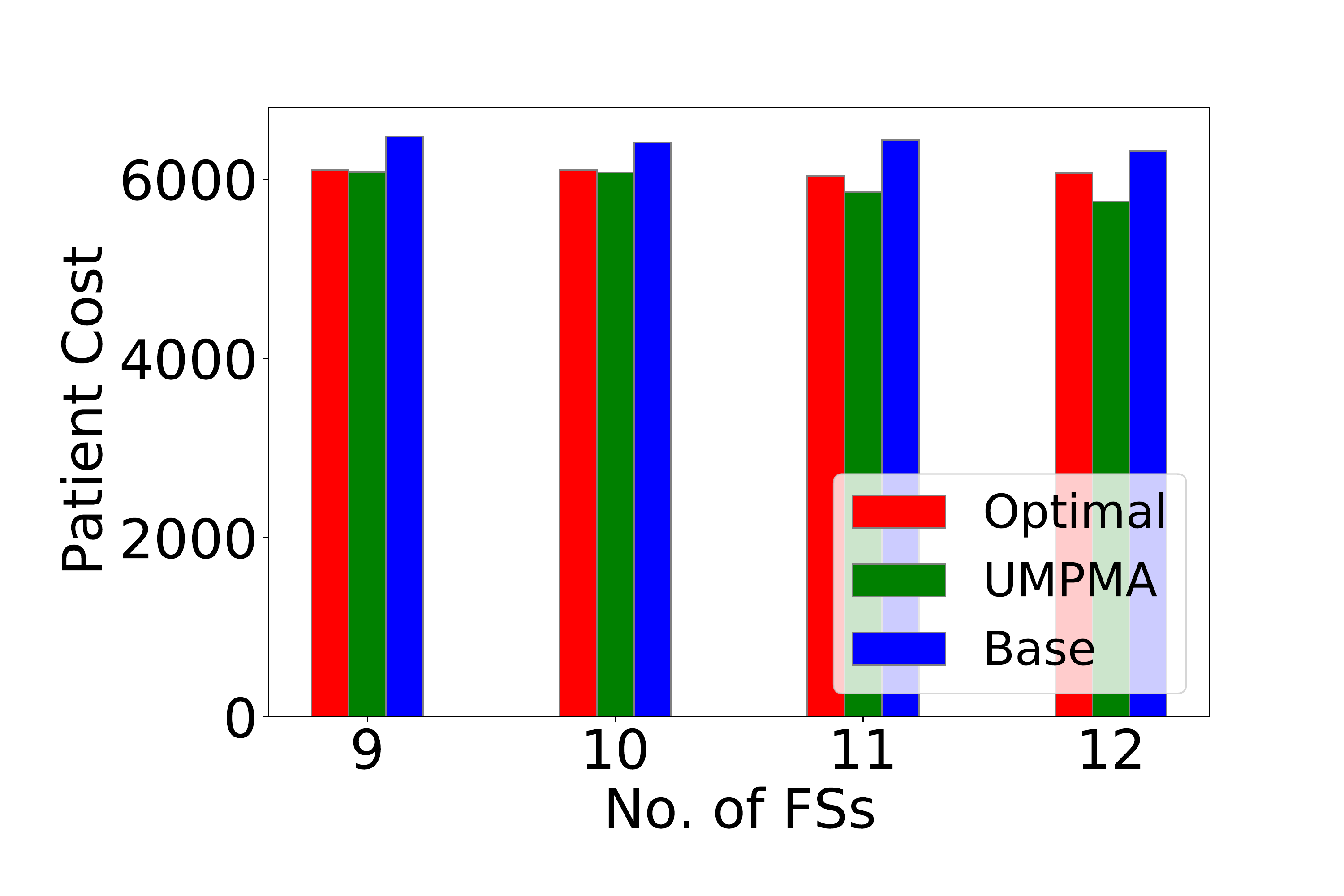}
 		\caption{No. of patients = 60.}
 		\label{plot_cost_60}
 	\end{subfigure}
 	\caption{Patients' cost comparison among different schemes.}
 	\label{plot_cost}
 		\vspace{-0.1in}
 \end{figure*}
 
To simulate the proposed health monitoring system, we have considered different number of patients and FSs in simulation environment. The size of the patient's data is randomly considered to be within 1 to 3 MB and the CPU cycles required are randomly considered to be between 100 to 1000 Megacycles \cite{ning2020mobile}. The value of $\delta$ is considered to be 250 ms. The bandwidth of a channel is taken as 5 MHz. Transmission power and noise are considered to be 0.1 Watts and -100 dBm, respectively as shown in Table \ref{sim_param}. Distance between any patient and FS is randomly taken as between 50 to 100 m. Channel gain is considered to be as $(dist_{p,f})^{-3}$ \cite{rappaport1996wireless}, where $dist_{p,f}$ is the distance between the LD and the FS. The computation capacity of the FS is taken as 22.4 GHz and that of LD is taken as 2.4 GHz. The patient criticalities are considered between 0 to 1 and are randomly generated. Although patient criticalities can be anything between 0 to $\infty$, it can be normalized between 0 to 1 for each patient. We are considering various parameters to evaluate the performance of our proposed algorithm. Python 3.9.0 platform is used to model the above simulation setup and execution of proposed heuristic.

IBM ILOG CPLEX Optimization Studio has been utilized to obtain optimal solutions for the comparison \cite{ilog}. The simulations were performed on a personal computer with processor Intel(R) Core(TM) i5-8250U CPU @ 1.60ghz.

Moreover, we propose another scheme, named \emph{Base}, to compare it with our proposed heuristic. Base scheme first considers patients that violates the constraint, sort them in decreasing order of their criticalities. Then, it allocates the patient one-by-one to an FS that results in increasing the utility by the maximum amount. It then considers, the remaining patients and order them by criticalities in decreasing order, and then repeats the same process of allocation.

\subsection{Simulation Results}
In this section, the results are presented on various aspects such as mentioned below:

\subsubsection{System Utility}

In Fig. \ref{plot_utility}, we have considered three cases, when the number of patients are 20, 40 and 60 and the number of FSs varies from 2 to 12. A general trend can be seen among all three cases. The proposed heuristic performs better than the Base scheme. The utility obtained by the proposed heuristic is $96\%$ of the optimal value compared to $56\%$ of that of Base scheme on an average. The reason is that the Base scheme allocates the patients in a particular order. Although the Base scheme is based on patient criticality which is an important factor for the system, however, it does not consider the data size and the CPU cycles of data packets. The proposed heuristic considers all the above factors to reach a sub-optimal utility within polynomial time complexity.

\subsubsection{Patient Cost}

In this section, we present simulation results for the patients' cost. From Fig. \ref{plot_cost}, it can be observed that the proposed heuristic generally results in lower patients' cost than the Base scheme. The reason is that the proposed heuristic considers different parameters and it allocates and re-allocates patients to maximise utility, resulting in lower patients' cost. However, the Base scheme does not re-allocate patients, resulting in higher patients' cost and lower utility. It can be seen that sometimes the patients' cost obtained by the proposed heuristic is lower than that obtained by the optimal solution. It is because optimal solution considers all possibilities and it could be a possibility that profit becomes a dominating factor in the optimal solution. However, our algorithm is more criticality-aware, thus it tries to increase the utility by lowering the patients' cost.

\begin{figure*}[ht!]
 	\centering
 	\begin{subfigure}[b]{0.325\textwidth}
 		\centering
 		\includegraphics[width=\textwidth]{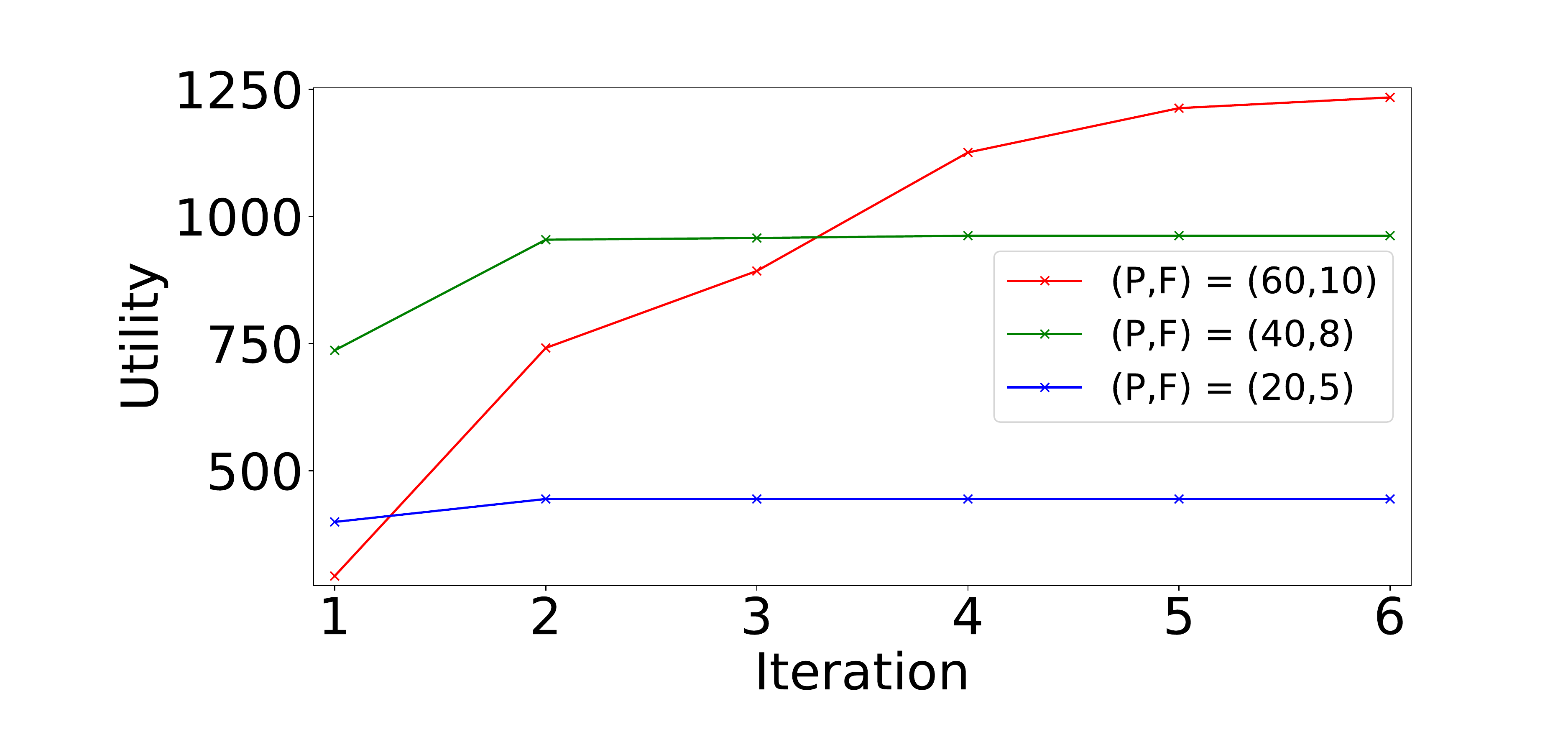}
 			\caption{Convergence Analysis.}
         	\label{plot_itr_converge}
 	\end{subfigure}
 	\begin{subfigure}[b]{0.325\textwidth}
 		\centering
 		\includegraphics[width=\textwidth]{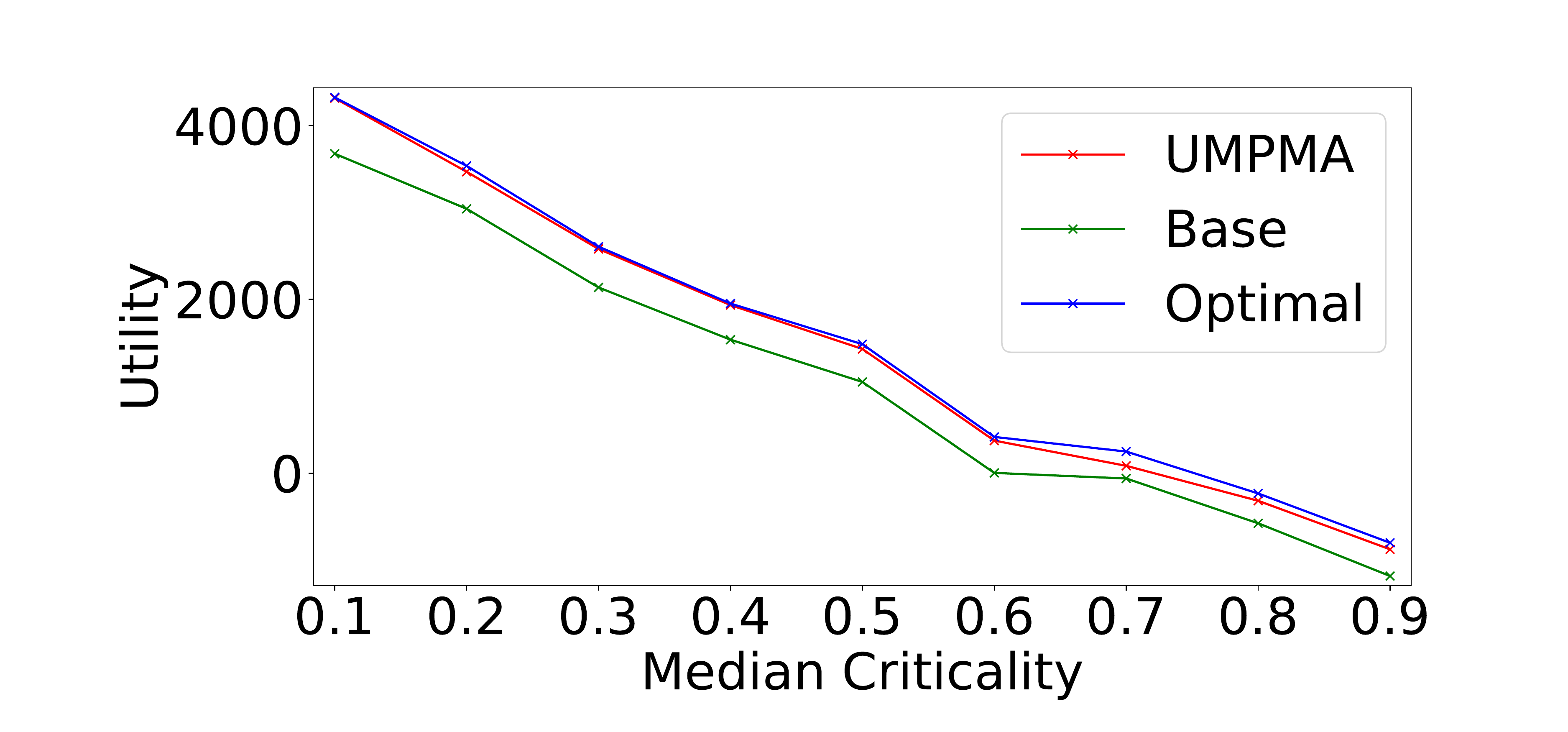}
 		\caption{Utility v/s Criticality}
	\label{plot_criticality_utility}
 	\end{subfigure}
 	\begin{subfigure}[b]{0.325\textwidth}
 		\centering
 		\includegraphics[width=\textwidth]{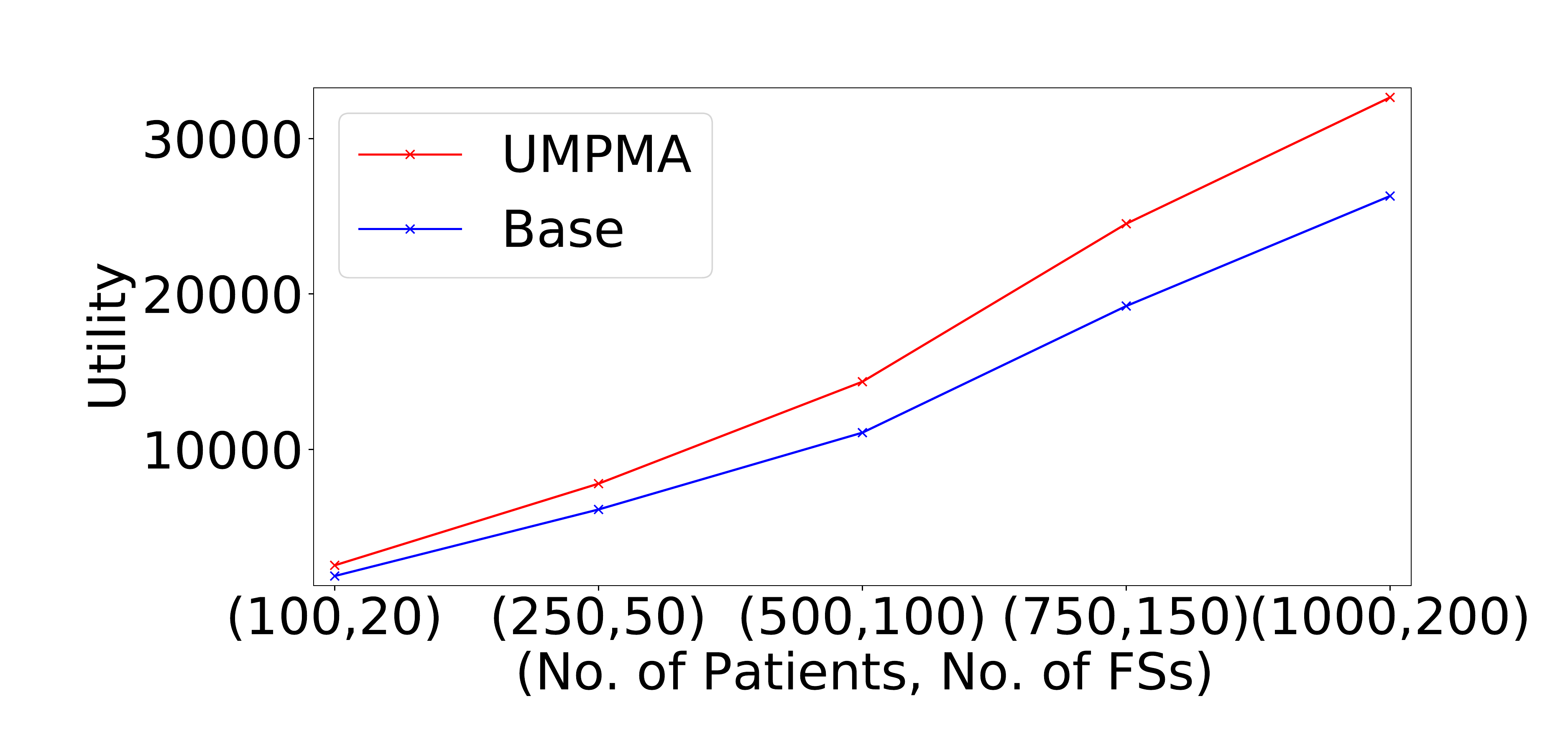}
 		\caption{Utility comparison for dense networks.}
    \label{plot_dense}
 	\end{subfigure}
 
 	\caption{Utility comparison against different parameters.}
 		\vspace{-0.1in}
 \end{figure*}

\subsubsection{Convergence Analysis}

Fig. \ref{plot_itr_converge} depicts the convergence of the algorithm for three different cases. From the result, we can observe that when $P = 60$, the algorithm converges in 6 iterations, and in 2 and 4 iterations for $P = 20$ and $P = 40$ respectively, on an average. As considered in the Section \ref{analysis}, the number of iterations are quite small and can be taken as a constant.

\subsubsection{Trade-off between utility and criticality}

We analyze the trade-off between utility and criticality through simulation. We have considered nine different median criticalities from 0.1 to 0.9. We have considered a fixed order of patients and then, assigned a median criticality to the middle patient. We randomly assign a criticality lower and higher than the median to the patients before and after the middle patient, respectively. In Fig. \ref{plot_criticality_utility}, it can be observed that as the median criticality increases, the system utility decreases. However, the relative trends remain the same and our proposed algorithm achieves a system utility of $97\%$ of the optimal whereas the Base scheme could only achieve $83\%$ of the optimal on an average.

\subsubsection{Utility comparison for dense networks}
We consider a large number of patients and FSs to analyse the proposed algorithm. However, due to the large number of patients and FSs, it is difficult to obtain an optimal solution using IBM ILOG CPLEX tool \cite{ilog}. Thus, we compare the results with the Base scheme only. In Fig. \ref{plot_dense}, we can see that the proposed algorithm gives better result than the base scheme. For a small number of patients and FSs, the utility gap between the proposed heuristic and the Base scheme is less, but as the number of patients and FSs increases, the gap also increases. Thus, the proposed heuristic is quite beneficial for dense networks. The Base scheme could only achieve a utility of $78\%$ of the utility achieved by the proposed heuristic.

\subsubsection{Execution time comparison}

\begin{figure}
    \centering
    \includegraphics[height=5cm, width=8.0 cm]{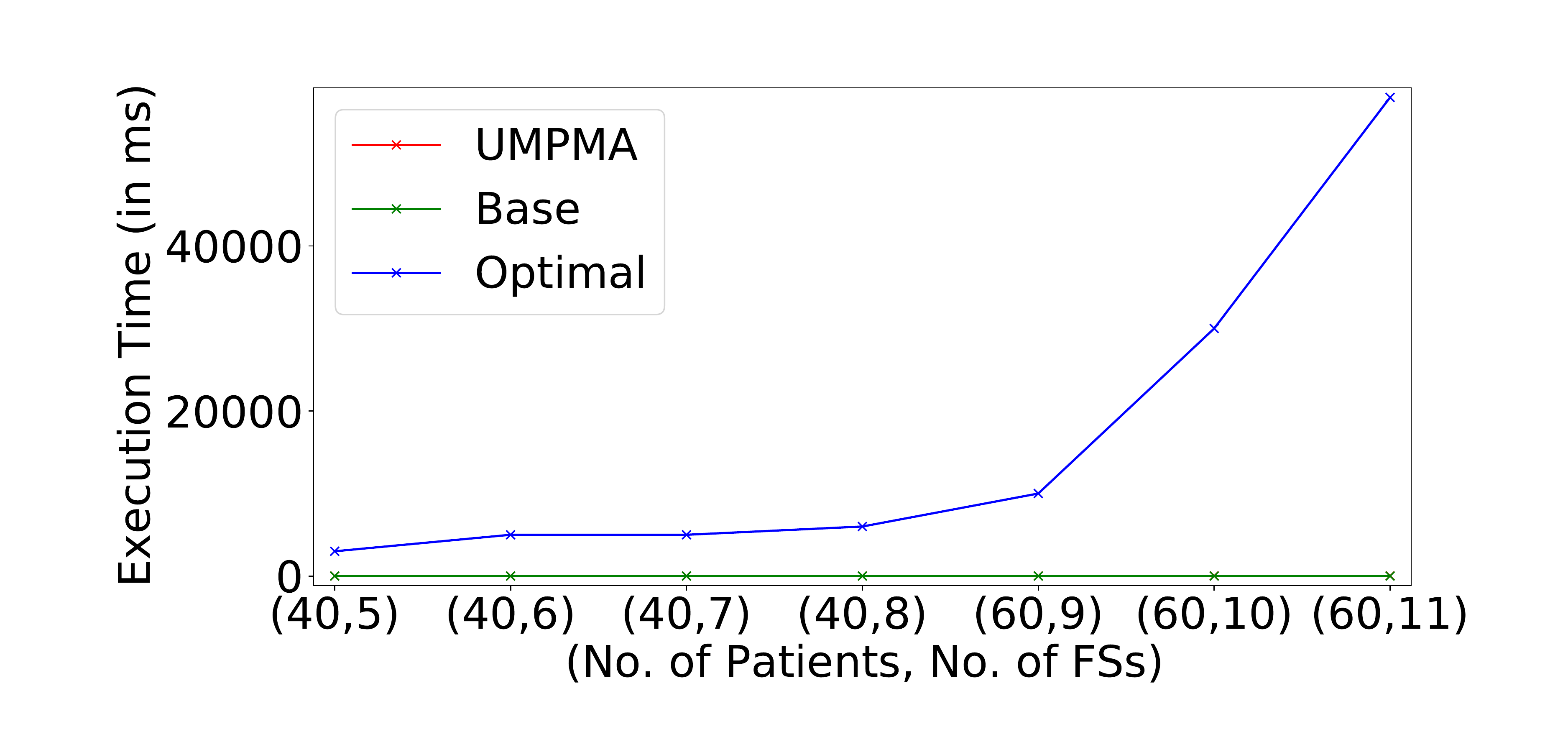}
    \caption{Execution time comparison}
    \label{plot_exec_time}
\end{figure}

In this section, we compare the execution time of three schemes. For Optimal, we consider the execution time as the time required by the ILOG CPLEX tool to reach the utility achieved by the proposed heuristic. All three schemes were run on the same machine and the execution time is taken as the average of execution times obtained by extensively executing these schemes. The proposed heuristic and the Base scheme complete their execution in a few milliseconds, thus, their plot is overlapping each other. However, the optimal takes some seconds to reach the utility achieved by the proposed heuristic. Therefore, we can say that the proposed heuristic achieves a sub-optimal utility in a small time compared to the optimal and its execution time is comparable to the Base scheme for smaller inputs. For more dense networks, the Base scheme is quite fast, but, the utility achieved is quite low. Thus, UMPMA emerges as a better scheme, balancing both the execution time and system utility.

\section{Conclusion}\label{Sec9}
In this paper, we have designed a beyond-WBAN based fog computing system for remote health monitoring. The main contribution was in the beyond-WBAN, where we formulated a problem based on the profit of medical center and the loss of patients, measured in terms of latency and criticalities. We then proposed a criticality-aware utility maximization heuristic (i.e., UMPMA) to maximize the utility in beyond-WBAN. The proposed heuristic is based on the swapping mechanism. Simulation results and evaluation of the UMPMA were presented to show the effectiveness of the proposed heuristic on various parameters. The proposed algorithm was demonstrated as criticality-aware, thus serving the purpose of the system. Through extensive simulations, we show that the proposed heuristic achieves an average utility of $96\%$ of the optimal, in polynomial time complexity.

This study leads to some future directions. The interference can be included in the model and thus, a sub-channel allocation problem can be considered with the current utility maximization problem. The role of doctors can be directly included in the system based on the criticality of the patients and it gives rise to a different pricing model as doctors are directly involved. Moreover, energy consumption is an important factor along-with the latency. Thus, a future study can be in the direction of energy-awareness together with the criticality-awareness. 

\noindent \textbf{Acknowledgments:} This work is partially supported by the Science and Engineering Research Board (SERB), Government of India under Grant SRG/2020/000318.

% \bibliographystyle{ieeetr}
% %\bibliography{bibl}
% \bibliography{IEEEabrv,bibl}
% %\bibliography{btp}

\begin{IEEEbiography}[{\includegraphics[width=1.1in,height=1.2in,clip,keepaspectratio]{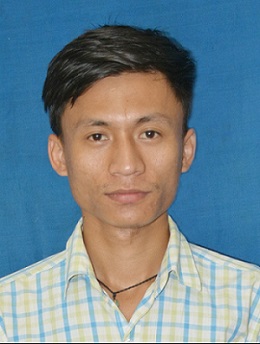}}]{Moirangthem Biken Singh} completed the B.Tech degree in Computer Science and Engineering from the National Institute of Technology Manipur, India, in 2018 and the M.Tech degree from National Institute of Technology Kurukshetra, India, in 2021. He is currently pursuing Ph.D. degree in Computer Science and Engineering, Indian Institute of Technology (BHU) Varanasi, India. His current research interest include AI, machine learning and FL in Smart Healthcare.
\end{IEEEbiography}

\begin{IEEEbiography}[{\includegraphics[width=1.1in,height=1.2in,clip,keepaspectratio]{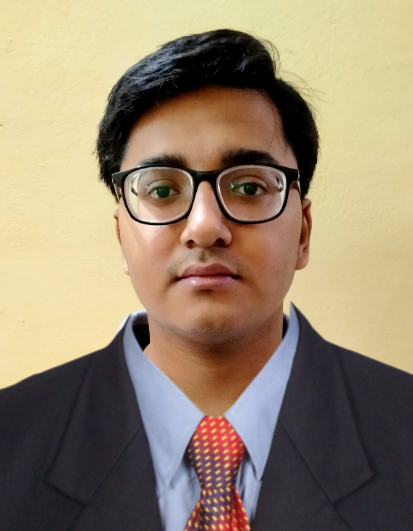}}]{Navneet Taunk} is an undergraduate student with the Department of Computer Science and Engineering, Indian Institute of Technology (BHU) Varanasi, India. His research interests include Design of Algorithms, Internet of Things (IoT) and Mathematical Modelling.
\end{IEEEbiography}

\begin{IEEEbiography}[{\includegraphics[width=1.1in,height=1.2in,clip,keepaspectratio]{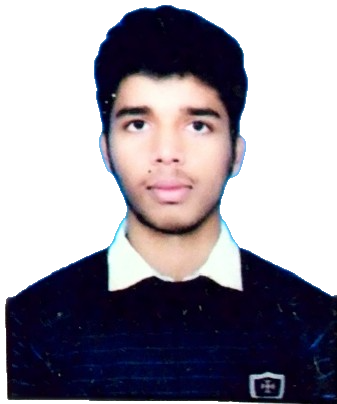}}]{Naveen Kumar Mall} is an undergraduate student with the Department of Computer Science and Engineering at the Indian Institute of Technology (BHU) Varanasi, India. His research interests include Deep learning, Machine learning and Internet of Things (IoT).
\end{IEEEbiography}

\begin{IEEEbiography}[{\includegraphics[width=1.1in,height=1.2in,clip,keepaspectratio]{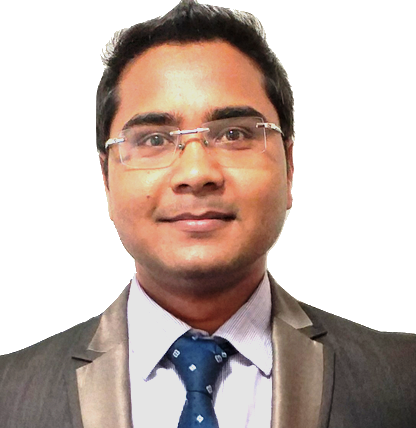}}]{Ajay Pratap} is an Assistant Professor with the Department of Computer Science and Engineering, Indian Institute of Technology (BHU) Varanasi, India. Before joining IIT (BHU), he was associated with the Department of Computer Science and Engineering, National Institute of Technology Karnataka (NITK) Surathkal, India, as an Assistant Professor from December 2019 to May 2020. He worked as a Postdoctoral Researcher in the Department of Computer Science at Missouri University of Science and Technology, USA, from August 2018 to December 2019. He completed his Ph.D. degree in Computer Science and Engineering from the Indian Institute of Technology Patna, India, in July 2018. His research interests include Cyber-Physical Systems, IoT-enabled Smart Environments, Mobile Computing and Networking, Statistical Learning, Algorithm Design for Next-generation Advanced Wireless Networks, Applied Graph Theory, and Game Theory. His current work is related to HetNet, Small Cells, Fog Computing, IoT, and D2D communication underlaying cellular 5G and beyond. His papers appeared in several international journals and conferences including IEEE Transactions on Mobile Computing, IEEE Transactions on Parallel and Distributed Systems, and IEEE LCN, etc. He has received several awards including the Best Paper Candidate Award and NSF travel grant for IEEE Smartcom'19 in the USA. 
\end{IEEEbiography}

\end{document}